\documentclass[jcp,floatfix, superscriptaddress,twocolumn,a4paper,nofootinbib]{revtex4-1}
\usepackage[T1]{fontenc}
\usepackage[utf8]{inputenc} % still required for pdflatex
\usepackage{textcomp}

\usepackage{graphicx}% Include figure files
\usepackage{amsmath}
\usepackage{xcolor}
    \usepackage{soul}

\newcommand{\ex}{\mathrm{ex}}
\newcommand{\eq}{\mathrm{eq}}
\newcommand{\inh}{\mathrm{inh}}
\newcommand{\out}{\mathrm{out}}
\newcommand{\act}{\mathrm{act}}
\newcommand{\eff}{\mathrm{eff}}
\newcommand{\bnd}{\mathrm{bound}}
\newcommand{\sat}{\mathrm{sat}}
\newcommand{\DoH}{\mathrm{DoH}}
\newcommand{\Da}{\mathrm{Da}}

\newcommand{\be}{\begin{equation}}
\newcommand{\ee}{\end{equation}}

\newcommand\ta[1]{\textcolor{black}{#1}}

\begin{document}

\title{How back reaction, hydrogen transport, \ta{and capillarity} control the performance of hydrogen release from liquid organic carriers}%trapping of bubbles inhibit hydrogen release from liquid organic carrier}}

%\title{
%High-order back reactions tune the performance of catalytic pellets\\
%Back reactions and transport control the performance of  LOHC dehydrogenation
%The unusual suspects in porous catalysts: back reactions and transport\\
%\ta{The unusual suspects in LOHC dehydrogenation: back reactions and transport}
%}

%\title{Confined catalysis: a tale of backreactions}
\author{T. Nizkaia} 
\affiliation{Helmholtz-Institute Erlangen-N\"urnberg for Renewable Energy (IET-2), Forschungszentrum J\"ulich, Cauerstraße 1, 91058 Erlangen, Germany}
\author{T. Solymosi} 
\affiliation{Helmholtz-Institute Erlangen-N\"urnberg for Renewable Energy (IET-2), Forschungszentrum J\"ulich, Cauerstraße 1, 91058 Erlangen, Germany}
\author{P. Malgaretti}
\email[Corresponding Author: ]{p.malgaretti@fz-juelich.de}
\affiliation{Helmholtz-Institute Erlangen-N\"urnberg for Renewable Energy (IET-2), Forschungszentrum J\"ulich, Cauerstraße 1, 91058 Erlangen, Germany}

\author{P. Wasserscheid}
\affiliation{Helmholtz-Institute Erlangen-N\"urnberg for Renewable Energy (IET-2), Forschungszentrum J\"ulich, Cauerstraße 1, 91058 Erlangen, Germany}
\affiliation{Lehrstuhl f\"ur Chemische Reaktionstechnik, Friedrich-Alexander-Universit\"at Erlangen-N\"urnberg, Egerlandstra\ss e 3, 91058 Erlangen, Germany}
\affiliation{Institut f\"ur nachhaltige Wasserstoffwirtschaft (INW), Forschungszentrum J\"ulich, Am Brainergy Park 4, 52428 Jülich, Germany}
\author{J. Harting}
\affiliation{Helmholtz-Institute Erlangen-N\"urnberg for Renewable Energy (IET-2), Forschungszentrum J\"ulich, Cauerstraße 1, 91058 Erlangen, Germany}
\affiliation{Department of Chemical and Biological Engineering and Department of Physics, Friedrich-Alexander-Universit\"at Erlangen-N\"urnberg, Cauerstraße 1, 91058 Erlangen, Germany} 

\date{\today}
\begin{abstract}
    \ta{We derive a theoretical model to elucidate the inhibition of catalytic activity during the dehydrogenation of Liquid Organic Hydrogen Carriers (LOHC). 
    Within our model, we account for the reversible nature of the hydrogenation–dehydrogenation reaction as well as the transport of both LOHC and produced hydrogen. 
    %Our analysis reveals that the transport of dissolved hydrogen, either via advection-diffusion or via bubbling, is the main limiting factor for the performance of porous catalysts. 
    %Moreover, we demonstrate that two distinct kinetic regimes — with high or strongly inhibited hydrogen production — can arise depending on whether dissolved hydrogen can be carried away 
    %Our analysis reveals that two distinct kinetic regimes — with high or strongly inhibited hydrogen production — can arise depending on whether dissolved hydrogen can be carried away. 
    %Moreover, we describe the dependence of the onset of bubbling on both the supersaturation of hydrogen and the capillary trapping of the bubbles within the porous catalyst. 
    %either via bubble breakthrough is enabled or suppressed \ta{by capillary trapping.} 
    Our analysis reveals that the main limiting factor for the performance of porous catalysts is the transport of dissolved hydrogen, which has been overlooked so far. In particular, we show that two distinct kinetic regimes can arise depending on whether hydrogen leaves the pellet in form of bubbles or via diffusion. Moreover, we derive the conditions for the onset of bubbling depending on hydrogen supersaturation and capillarity.} 
    Beyond LOHC systems, our findings are applicable to a broader class of reversible reactions, particularly those involving volatile products that can leave the liquid reaction medium in the form of bubbles.\\
    
\end{abstract}


\maketitle
\section{Introduction}
Hydrogen has emerged as a promising candidate for clean energy storage and transport, playing a key role in the transition towards a sustainable energy future~\cite{staffell_role_2019,singla_hydrogen_2021,GreenH2Review2023}. However, the practical challenges in hydrogen storage and transportation remain significant: since hydrogen is a highly flammable gas, its storage in large quantities requires expert knowledge and dedicated infrastructures. A potential solution lies in the use of Liquid Organic Hydrogen Carriers (LOHCs): compounds that can chemically bind and release hydrogen in a reversible way~\cite{taube_system_1983,newson_seasonal_1998,teichmann2011future,heublein_hydrogen_2020,rao2020potential,sekine_recent_2021,jo_recent_2022}. By this method, hydrogen is chemically bonded to the LOHC molecules during the hydrogenation reaction, allowing for safe and efficient storage at ambient conditions. When needed, hydrogen can be released from the LOHC molecules using the dehydrogenation reaction and utilized for various applications. Both hydrogenation and dehydrogenation of LOHC compounds require the use of a catalyst, usually in the form of metallic nanoparticles embedded in a porous support (catalytic pellet) ~\cite{salman2022catalysis,KWAK2021114124}. 
\begin{figure}[h!]
\centering
    \includegraphics[width=0.9\columnwidth]{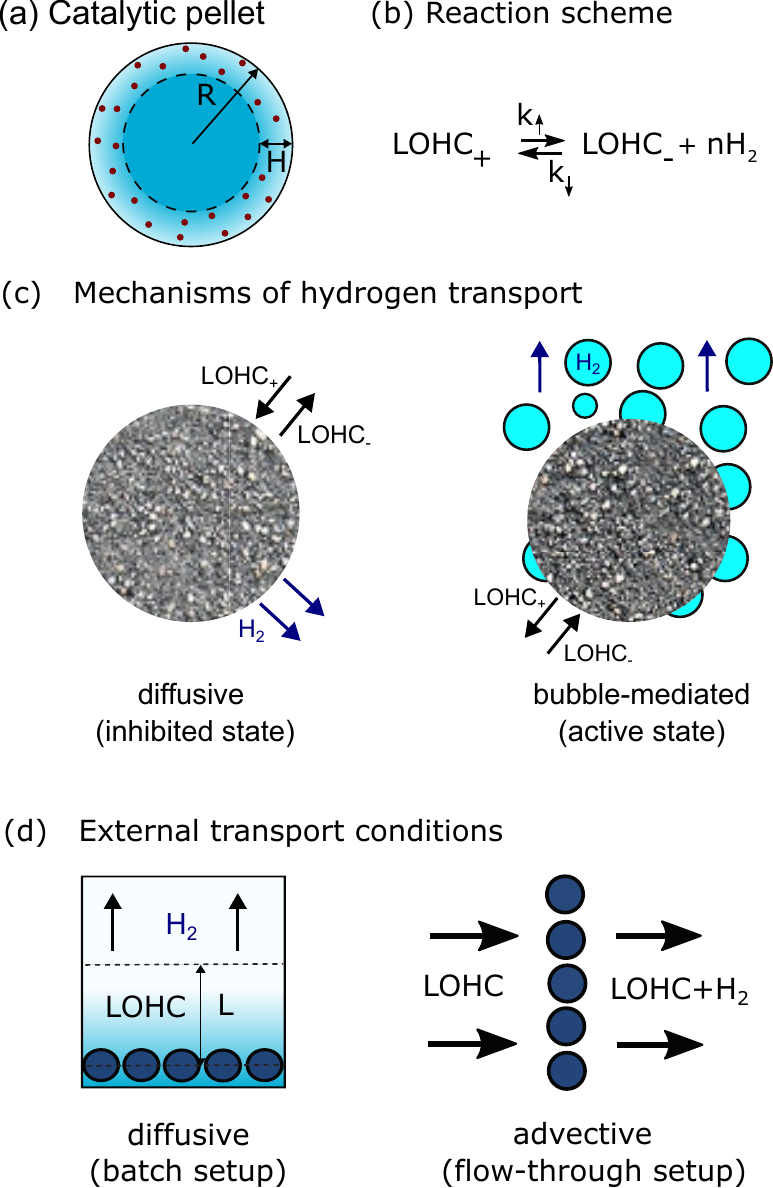}
\caption{(a) Sketch of a catalytic pellet with the active sites, e.g. suitable surface atoms of the supported metallic nanoparticles, being
schematically shown in red. These are homogeneously distributed in the active layer of thickness H of an egg-shell catalyst pellet. Dissolved hydrogen is indicated as shades of blue. (b)~Simplified scheme of reversible dehydrogenation reaction. (c)~Simplified model of a batch dehydrogenation experiment (distribution of dissolved hydrogen shown in blue). (d)~Simplified model of the inlet region of a flow-through reactor.}\label{fig:sketch}
\end{figure}

The overall performance of a catalytic pellet is determined by the interplay of the intrinsic kinetics at the catalyst active sites and the transport of reactants and products. %The influence of transport is usually measured in terms of efficiency factor of the pellet: the ratio of the flux of products from a real pellet compared with the hypothetical case in which reactants and products are distributed in the pellet homogeneously~\cite{huang1965mathematical,carberry1962micro,rios2023revisiting}. 
Transport phenomena are particularly important for easily reversible reactions, for which accumulation of products shifts the chemical equilibrium locally and can slow down the reaction or stop it completely~\cite{kao1968effectiveness,rios2023revisiting}. This is indeed the case for hydrogenation/dehydrogenation of LOHC compounds: at the same temperature, both hydrogenation and dehydrogenation may occur depending on its current degree of hydrogenation of the LOHC mixture and on the partial pressure of hydrogen in contact with the liquid~\cite{durr2021experimental,jorschick2017hydrogen}. Therefore, we can expect the local concentration of dissolved hydrogen to have a significant impact on the efficiency of hydrogen production.

In recent years, significant advances have been made in elucidating the mechanisms of catalytic hydrogenation/dehydrogenation~\cite{taube_system_1983,newson_seasonal_1998,teichmann2011future,heublein_hydrogen_2020,rao2020potential,sekine_recent_2021,jo_recent_2022,rakic2023liquid} and in modeling these processes on the reactor scale~\cite{van2024intensified}. While the importance of back reaction and its dependence on the operational pressure has been recognized~\cite{geisselbrecht2024modeling, fan2025continuous,kadar2024boosting,gambini2024flow}, the influence of hydrogen transport on dehydrogenation rate has not been considered so far: it is generally assumed that \ta{hydrogen leaves the system immediately in form of bubbles and its concentration in the reaction mixture is always at equilibrium with the gas phase. However, this is not always the case.} 

\ta{It has been observed that at the same thermodynamic conditions, catalytic pellets used for LOHC dehydrogenation can be either in an active state (intense bubble formation at the surface of the pellet, high productivity) or in an inhibited state (no bubbling, low productivity), see Fig.~\ref{fig:sketch}. This phenomenon of pellet inhibition has been studied recently in two experimental setups: small batch~\cite{solymosi2022nucleation} and flow-through~\cite{uhrig2024reactivation}, with markedly different results. While in batch experiments the inhibited pellets produced 50 times less hydrogen than active ones~\cite{solymosi2022nucleation}, in a flow-through reactor the inhibited pellets performed surprisingly well, with productivity reduction of just $10-20\%$~\cite{uhrig2024reactivation}. The transition between the states was attributed to bubble nucleation in the pores of the pellet and different reactivation strategies have been proposed to avoid the undesirable inhibited state \cite{solymosi2022nucleation}. However, the physical mechanism inhibiting hydrogen production in absence of bubbling and the observed sensitivity to external transport conditions remained unexplained.}  

\ta{In this paper we demonstrate that the inhibited state can result from the interplay between the reversible hydrogenation/dehydrogenation reaction, hydrogen transport and capillary trapping of gas bubbles in the porous space of the pellet. To do so, we develop a reaction--diffusion model that allows us to calculate hydrogen production and supersaturation in a porous pellet depending on the hydrogen concentration in the surrounding liquid.  %for reversible hydrogenation/dehydrogenation reactions with a non-linear dependency of the hydrogenation rate on
First, we explore the dependency of hydrogen production on transport conditions inside and outside of the pellet. We show that slow transport of produced hydrogen in non-bubbling regime leads accumulation of hydrogen at the active sites and inhibition of dehydrogenation due to the back reaction. This explains both drastic reduction of hydrogen production in the batch experiment and the improved performance of inhibited pellets in a flow-through setup. Using available data on physical properties of the H0-DBT/H18-DBT LOHC system ~\cite{durr2021experimental,aslam2016measurement,heller2016binary} we obtain quantitative estimates for the degree of inhibition in batch and flow through experiments and find them in good agreement with observations \cite{solymosi2022nucleation,uhrig2024reactivation}. Finally, we estimate the supersaturation of hydrogen inside the catalytic pellets in the batch experiment and show that it is consistent with the hypothesis of capillary trapping preventing activation. }
 
%In particular, we show that hydrogen transport plays a crucial role in deactivation of catalytic pellets, observed for some LOHC systems \cite{solymosi2022nucleation}. 

%Our results show that the different reduction in performance for nucleation-inhibited pellets observed in flow-through~\cite{uhrig2024reactivation} or batch~\cite{solymosi2022nucleation} experiments on H18-DBT (perhydro-dibenzyltoluene) dehydrogenation is due to two typically overlooked phenomena: the back reaction with dissolved hydrogen and \ta{capillary trapping of gas bubbles in the porous space of the catalyst.} 

%Accordingly, parameters that have been typically disregarded, such as the distance between the pellets and the liquid-gas interface in batch experiments, are indeed crucial in determining the performance of the pellets. 
%Finally, we use our model to discuss the dependence of the performance of the pellets on their inner structure. 

\section{Reaction--diffusion model in the pellet}
To estimate the hydrogen production of a catalytic pellet in the absence of bubbling, we need to model the reaction and transport of hydrogen and LOHC compounds inside the pellet. Dehydrogenation of most LOHC compounds proceeds in several steps, producing a number of long-lived partially dehydrogenated intermediates LOHC-X, where X denotes the number of hydrogen atoms bound to LOHC~\cite{sekine_recent_2021}.
The local composition of the mixture of LOHC compounds is defined by a system of reaction--diffusion equations as~\cite{wakao1964diffusion, huang1965mathematical}
\begin{eqnarray}
\frac{\partial C_X}{\partial t}=\nabla\left[D^\eff_X\nabla C_X\right]+S_X\cdot \phi(\mathbf x),\label{eq:diffusion}
\end{eqnarray}
where $C_X(\mathbf{x})$ are molar concentrations of partially hydrogenated LOHC compounds, $D^{\eff}_{X}$ are effective diffusion coefficients of LOHC compounds in a porous medium, the source terms $S_{X}$ are the local rates of production/consumption of partially dehydrogenated LOHC compounds due to hydrogenation/dehydrogenation and $\phi(\mathbf{x})$ is a non-dimensional function describing the spatial distribution of the available catalyst surface in the pellet.
The source of hydrogen is then defined as
\begin{equation}
S_{H_2}
=-\dfrac{1}{2}\sum\limits_X X S_X. \label{eq:h2balance} 
\end{equation}

Tracking the reaction intermediates LOHC-X is important for understanding the hydrogenation and dehydrogenation pathways~\cite{do2016hydrogenation, shi2023dehydrogenation}. However, these intermediates are difficult to measure in situ, and the detailed composition of the mixture is typically not considered in the literature.  Instead, the composition of the mixture of LOHC compounds is characterised by a degree of hydrogenation (DoH)~\cite{aslam2018thermophysical,bioucas2020thermal,durr2021experimental,park2021experimental}
\be
\DoH=\dfrac{C_{H_2}^{\bnd}}{n \bar C},
\label{eq:def_c+}
\ee 
which is the ratio of the number of $H_2$ molecules bound to LOHC molecules (per unit volume)
\begin{align}
 C_{H_2}^{\bnd}=\sum\limits_X \dfrac{X}{2}\cdot C_X   
\end{align}
to the loading capacity of the LOHC system: 
\begin{align}
n\bar C=n\sum\limits_X C_X,
\end{align} 
where $\bar{C}$ is the molar density of the LOHC mixture
and $n$ is the maximum number of $H_2$ molecules that can be chemically incorporated into an LOHC molecule. 
\iffalse
Typical values of $n$ are:
\begin{eqnarray*}
    n=4\;\text{(N-Ethylcarbazole)},\\
    n=6\;\text{(H12-DPM, H12-BT)},\\
    n=9\;\text{(H18-DBT).}
\end{eqnarray*}
\fi
%Thus $c_+=1$ corresponds to fully hydrogenated LOHC and $c_+=0$ to fully dehydrogenated one.

\ta{In the literature, the physical properties of LOHC systems and the data on the reaction equilibrium are usually available as a function of DoH. Therefore, we aim to construct the reaction-diffusion model based directly on the local degree of hydrogenation.
This approach becomes exact when dehydrogenation occurs via a single pathway: then each value of DoH corresponds to a certain composition of the LOHC mixture and DoH is enough to describe the system.  }

\ta{To do so we note that at} working temperatures ($500-600 \,\mathrm{K}$), the differences in molar densities and binary diffusion coefficients of typical LOHC compounds in hydrogenated and dehydrogenated forms do not exceed $10\%$~\cite{heller2016binary,kerscher2020thermophysical}. We therefore assume that the molar density of the LOHC mixture is composition-independent ($\bar{C}= const$), and the
diffusion coefficients of all partially dehydrogenated LOHC compounds are equal and also do not depend on the mixture composition:
\begin{equation}
D_X= D_{+
}=const\,
\label{eq:D_const} 
\end{equation}
This allows us to write a reaction--diffusion model for an LOHC mixture with varying degree of hydrogenation without resolving for intermediates. Indeed, we can multiply the reaction--diffusion equations Eqs.~\eqref{eq:diffusion} by $X/2$ (the number of $H_2$ molecules, 'stored' in one LOHC-X molecule), sum them up and obtain a reaction--diffusion equation for bound hydrogen
\begin{align}
    \dfrac{\partial C_{H_2}^\bnd}{\partial t}=D_+^\eff \nabla^2 C_{H_2}^\bnd-S_{H_2}\,, \label{eq:ch2_bound}
\end{align}
which is complemented by a reaction--diffusion equation for the concentration $C_{H_2}(\mathbf{x})$ of dissolved hydrogen:
\begin{align}
    \dfrac{\partial C_{H_2}}{\partial t}=D_{H_2}^\eff \nabla^2 C_{H_2}+S_{H_2} \label{eq:ch2_free}
\end{align}
The source term $S_{H_2}$ describes the reversible chemical binding of $H_2$ to the mixture of LOHC molecules. We represent the reaction as a combination of forward (dehydrogenation) and back (hydrogenation) reactions happening simultaneously at different rates (see Fig.~\ref{fig:sketch}(b)):
\begin{align}
S_{H_2}=k_{\uparrow}C_{H_2}^\bnd -k_{\downarrow}(n\bar C-C_{H_2}^{\bnd})
\end{align}
Here, $k_{\uparrow},k_{\downarrow}$ are the effective binding/unbinding rates and ($n\bar C-C_{H_2}^\bnd$) is the number of free ``binding slots'' in the mixture of LOHC compounds.
\ta{The source term, $S_{H_2}$, results from the combination of  the effective hydrogen production, $k_\uparrow C_{H_2}^\bnd$, and consumption via rehydrogenation, $k_\downarrow C_{H_2}^\bnd(n\bar C-C_{H_2}^{\bnd})$, by all intermediate steps. The effective parameters $k_{\uparrow,\downarrow}$ should in principle depend on $\DoH$, however, in order to keep the model as simple as possible, here we disregard such a dependence.}

Unlike dehydrogenation, the hydrogenation reaction involves dissolved hydrogen; therefore its rate depends on the $H_2$ concentration near catalytic sites. Using the data in Ref.~\cite{durr2021experimental}, we approximate the empirical dependency of the hydrogenation rate on hydrogen concentration with a power law (see Appendix~\ref{sec:params}) as
\begin{align}
k_{\uparrow}=const,\hspace{1cm} k_{\downarrow}= k_{\uparrow}\cdot \gamma \cdot \left(\dfrac{C_{H_2}}{C_{H_2}^{\sat}}\right)^{m},
\label{eq:k-up-down}
\end{align}
where $m$ is the effective reaction order with respect to hydrogen, $C_{H_2}^{\sat}$ is the saturation concentration of $H_2$ in the LOHC mixture and $\gamma$ is the equilibrium constant of hydrogenation (both taken at atmospheric pressure) given by
\begin{align}
    \gamma=\dfrac{\DoH^{\eq}(P_a)}{1-\DoH^{\eq}(P_a)},
\label{eq:def_gamma}
\end{align}
where $\DoH^{\eq}(P_a)$ is the equilibrium degree of hydrogenation at fixed working temperature and atmospheric pressure $P_a$. We recall that since hydrogenation is in reality a multistep reaction, its effective reaction order can be different from unity. 

%Reaction diffusion equations Eqs.\eqref{eq:c}
%\begin{equation}
%    C_{H_2}^{\bnd}(R)=n\bar{C}\DoH^{\ex},\;C_{H_2}(R)=n\bar{C}\DoH^{\ex},
%\end{equation}

We consider a spherical pellet of radius $R$ and assume that all distributions are spherically symmetric, i.e. depend only on the spherical coordinate $\bar{r}$.
%\ta{The amount of hydrogen produced by the pellet per unit time can be calculated by integrating the diffusive flux of dissolved $H_2$ leaving the pellet:}
%\begin{equation}
%    Q_{H_2}=-4\pi R^2D_{H_2}^{\eff}\left.\frac{d C_{H_2}}{d\bar{r}} \right|_{\bar{r}=R}\label{eq:Qinh_dim}
  %  =-4\pi RD_{H_2}^{\eff}C_{H_2}^{\sat}\left.\frac{\partial c_{H_2}}{\partial r}\right|_{r=1},\label{eq:dim_flux}
%\end{equation}
%\ta{or, alternatively, the same value can be obtained by calculating the amount of LOHC-bound hydrogen entering the pellet:}
The hydrogen production rate of the pellet can be determined by integrating the flux of LOHC-bound hydrogen entering the pellet over its surface:
\begin{equation}
Q_{H_2}=4\pi R^2D_{+}^{\eff}\left.\frac{d C_{H_2}^\bnd}{d\bar{r}} \right|_{\bar{r}=R}.
\end{equation}
Note that this definition of $Q_{H_2}$ is valid for both the nucleation-inhibited (non-bubbling) state (when it is equal to the diffusive flow of produced $H_2$) and the active state (when hydrogen leaves the system in the form of bubbles and the diffusive flux of $H_2$ is undefined). 

Taking the saturation concentration as a reference value for the amount of dissolved hydrogen, we define the following non-dimensional variables:
\begin{eqnarray}
   r=\dfrac{\bar{r}}{R},\;c_+=\dfrac{C_{H_2}^{\bnd}}{n\bar{C}},\;c_{H_2}=\dfrac{C_{H_2}}{C_{H_2}^{\sat}},\;s_{H_2}=\dfrac{S_{H_2}}{k_{\uparrow}n\bar{C}}\label{eq:def_nondim}
   \end{eqnarray}
Substituting them into Eqns.~\eqref{eq:ch2_bound},\eqref{eq:ch2_free}, gives
\begin{eqnarray}
\nabla^2 c_+-\Da_+  s_{H_2}(c_+,c_{H_2})\phi(r)=0, \label{eq:cp_nd}\\
\nabla^2 c_{H_2}+\dfrac{\bar{C}}{C_{H_2}^{\sat}}\Da_{H_2}s_{H_2}(c_+,c_{H_2})\phi(r)=0, \label{eq:ch2_nd}
\end{eqnarray}
where 
\begin{equation}
s_{H_2}(c_+,c_{H_2})=c_+-\gamma\cdot \left(c_{H_2}\right)^m\cdot (1-c_+)
\label{eq:kinetics}
\end{equation}
is the non-dimensional reaction term obtained using Eq.~\eqref{eq:k-up-down} and
\begin{equation}
\Da_+=\dfrac{k_{\uparrow} R^2}{D_{+}^\eff}, \; \Da_{H_2}=\dfrac{n\cdot k_{\uparrow}R^2}{D_{H_2}^\eff} \label{eq:Da_def}
\end{equation}
are the Damk\"ohler numbers for the LOHC compounds and hydrogen, respectively. They are defined as the ratio of reaction to diffusion rates. We recall that the factor $n$ in Eq.~\eqref{eq:Da_def} accounts for the fact that full dehydrogenation of one LOHC molecule produces $n$ molecules of $H_2$, and the ratio $\dfrac{\bar C}{C_{H_2}^{\sat}}$ appears because we use different reference concentrations for the LOHC mixture and $H_2$ in Eqs.~\eqref{eq:def_nondim}.
Finally, Eqs.~\eqref{eq:cp_nd} and \eqref{eq:ch2_nd} are complemented with no-flux boundary conditions at the center of the pellet
and continuity conditions at the boundary of the pellet:
\begin{align}
  c_+(r=1)=c_+^{\ex},\quad c_{H_2}(r=1)=c_{H_2}^{\ex}.\label{eq:bc1}
\end{align}
Here, $c_+^{\ex}$ and $c_{H_2}^{\ex}$ are the local degree hydrogenation and the hydrogen supersaturation at the external surface of the pellet. 

Then, the non-dimensional hydrogen flux in the inhibited state (''inh'') can be introduced as follows:
\begin{eqnarray}
    q_{H_2}^{\inh}(c_{H_2}^{\ex},c_{+}^{\ex})=\dfrac{Q_{H_2}^{\inh}}{4\pi R D_{+}^{\eff}n\bar{C}}=\left.\dfrac{\partial c^{\inh}_{+}}{\partial r}\right|_{r=1}%=\ta{-\epsilon\left.\dfrac{\partial c_{H_2}}{\partial r}\right|_{r=1}}
    \label{eq:qinh_nondim}
\end{eqnarray}
Here, $c^{\inh}_+$ is the solution of Eqs.~\eqref{eq:cp_nd} and \eqref{eq:ch2_nd} with boundary conditions given by Eq.~\eqref{eq:bc1}. 

In the active state (``act''), when hydrogen is taken away by bubbles very efficiently, we can assume $C_{H_2} = C_{H_2}^{\sat}$ (no supersaturation) throughout the pellet. %The amount of gaseous hydrogen produced by the pellet will be equal to the amount of LOHC-bound hydrogen consumed. 
The flux of hydrogen in the active state then reads:
\begin{eqnarray}
    q_{H_2}^{\act}(c_{+}^{\ex})=\dfrac{Q_{H_2}^{\act}}{4\pi R D_{+}^{\eff}n\bar{C}}=\left.\dfrac{\partial c^{\act}_{+}}{\partial r}\right|_{r=1},\label{eq:qact_nondim}
\end{eqnarray}
where $c^{\act}_{+}$ is the solution of Eq.~\eqref{eq:cp_nd} with $c_{H_2}\equiv 1$ and the boundary condition given by Eq.~\eqref{eq:bc1}(left).

\ta{The active state is the 'standard' operating state of a catalytic pellet and its productivity does not depend on the external conditions of hydrogen transport. Therefore, we will use it as reference state and will focus on the ratio, $q_{H_2}^{\inh}/q_{H_2}^{\act}$, of hydrogen production in the inhibited and the active state. As we shall see later, this ratio can vary tremendously. Experimentally, it changes from $q_{H_2}^{\inh}/q_{H_2}^{\act}\simeq 1/50$~\cite{solymosi2022nucleation} for batch experiments to  $q_{H_2}^{\inh}/q_{H_2}^{\act}\simeq 0.8$~\cite{uhrig2024reactivation} for flow-through experiments.}  

\begin{figure}[t]
\centering
\includegraphics[width=0.8\columnwidth]{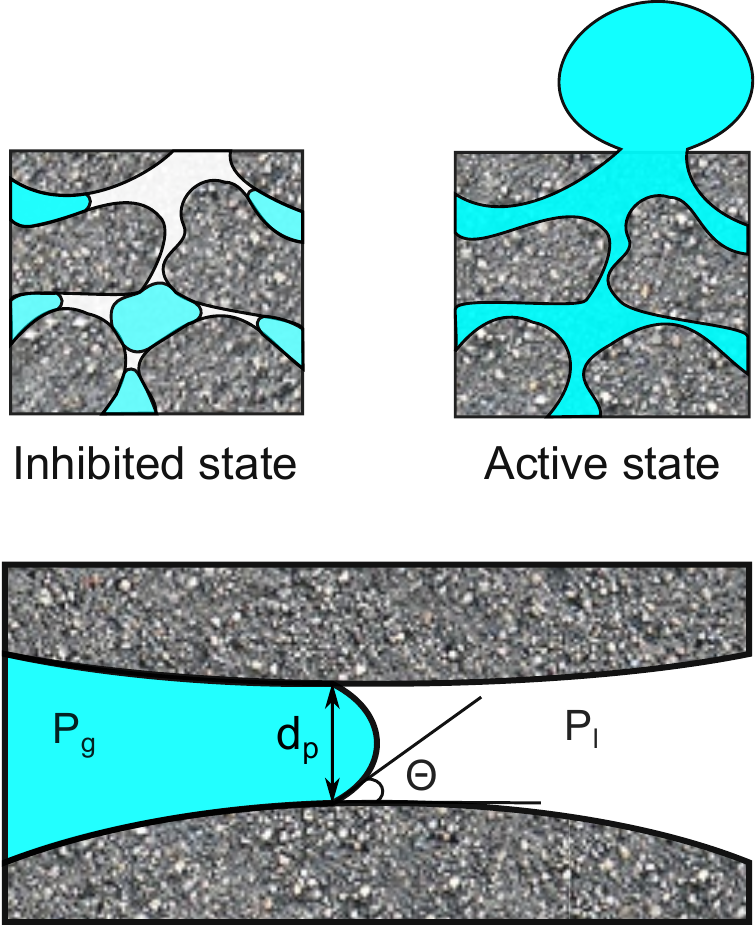}
\caption{A sketch of capillary trapped bubbles (top,left) and percolating bubble cluster (top, right) and an illustration of breakthrough conditions for a single bottleneck (bottom).}\label{fig:activation}
\end{figure}
\section{Results and discussion}

Hydrogen production in nucleation-inhibited pellets %and their transition into the active state \
depends on a complex interplay of reaction kinetics, transport properties, and external conditions. %In this section we use the reaction-diffusion model to analyze hydrogen flux under non-bubbling (inhibited) conditions to highlight the critical role of hydrogen transport outside the pellet. %
Accordingly, here we restrict ourselves to the case of a spherical catalytic pellet
with radius $R$, where the active catalytic sites are uniformly distributed in an outer shell of thickness $H$ (see Fig.~\ref{fig:sketch}(a)). The distribution function in Eqs.~\eqref{eq:cp_nd} and \eqref{eq:ch2_nd} then takes the form
\begin{equation}
    \phi(r)=\left\{
    \begin{array}{ll}
    0,&r<1-h\\
    1,&r\geq1-h,
    \end{array}\right.
\end{equation}
with $h=H/R$. 
In order to compare against experimental data, it is useful to introduce here a new non-dimensional parameter
\begin{equation}    \epsilon=\dfrac{D_+C_{H_2}^{\sat}}{D_{H_2}n\bar{C}},   \label{eq:eps}
\end{equation}
which allows to rewrite Eq.~\eqref{eq:ch2_nd} in a more compact form: 
\begin{equation}
    \nabla^2 c_{H_2}+\epsilon^{-1}\Da_{+}S_{H_2}(c_+,c_{H_2})\phi(r)=0 \label{eq:ch2_eps}
\end{equation}
%\ta{For typical LOHCs at the reaction temperature we have the ratio of diffusion coefficients $\dfrac{D_{H_2}}{D_{+}}=20\ldots30$ and hydrogen solubility $\dfrac{C_{H_2}^{\sat}}{\bar{C}}=4-8\cdot 10^{-3}$~\cite{heller2016binary,aslam2016measurement}, so that $\epsilon\approx 10^{-2}\ll1$.}
To allow for the direct comparison with experiment we use (unless stated otherwise) the following set of non-dimensional parameters, calculated using for H18-DBT at $T=573 \,\mathrm{K}$ using literature data \cite{heller2016binary,bioucas2020thermal,durr2021experimental}
(see Appendix~\ref{sec:params}):
\begin{equation}
\gamma=0.087, \; m=1.6, \;\epsilon=0.0125, \;h=0.1.\label{eq:default_pars}
\end{equation}
The Dammk\"ohler number $\Da_+$ typically varies between 10 and 100 (see Appendix~\ref{sec:params}).
%
%\textbf{recalculated with new definition of DaH2}:
%\begin{equation}
%\Da_+=10..100.
%%\Da_{H_2}=10^2..10^3
%%\Da_{H_2}=0.5..5
%%\Da_{H_2}=5..100 %with n
%\label{eq:default_Da}
%\end{equation}
%We are therefore most interested in the behavior of the system in diffusion-limited case, which corresponds to large $\Da_{H_2}$.
\subsection{Hydrogen production depending on external conditions: $c_{H_2}^{\ex}$ and $c_+^{\ex}$}

From experimental data, it is known that, in contrast to active pellets, nucleation-inhibited pellets exhibit strongly condition-dependent productivity. Their performance ranges from approximately 
$80–-90$\% of the active state in a flow-through reactor~\cite{uhrig2024reactivation} to less than 2\% in batch experiments~\cite{solymosi2022nucleation}. This pronounced change in the performance of inhibited pellets when transitioning from batch to flow-through conditions is striking, as the experiments were conducted under very similar thermodynamic conditions. Here, we show that this behavior can be attributed to differences in the hydrogen concentration at the pellet surface, arising from distinct transport conditions.
\begin{figure}[h!]
\centering
\includegraphics[width=0.98\columnwidth]{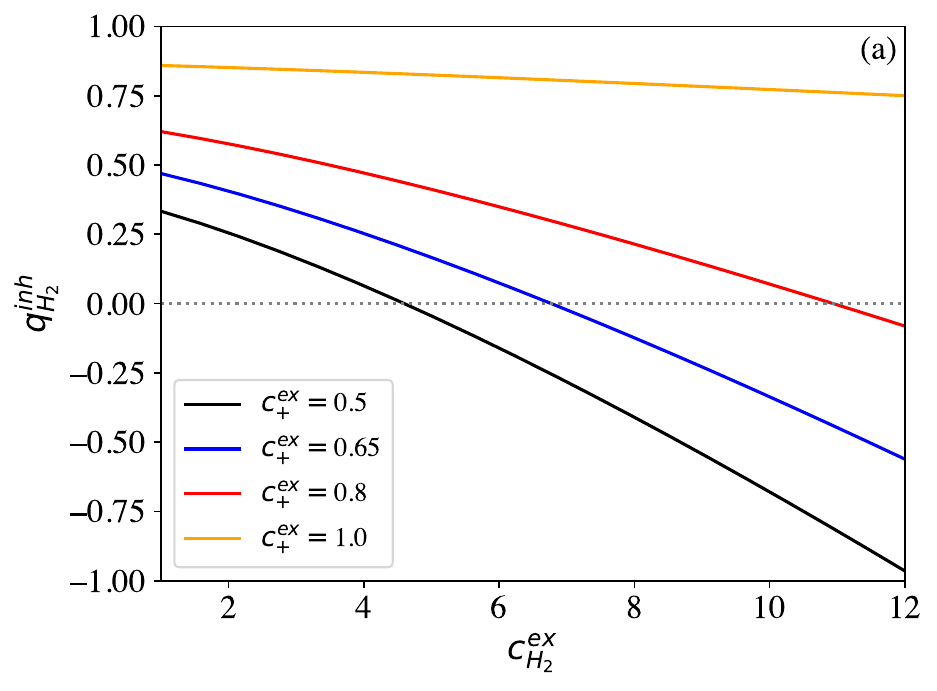}
\includegraphics[width=0.98\columnwidth]{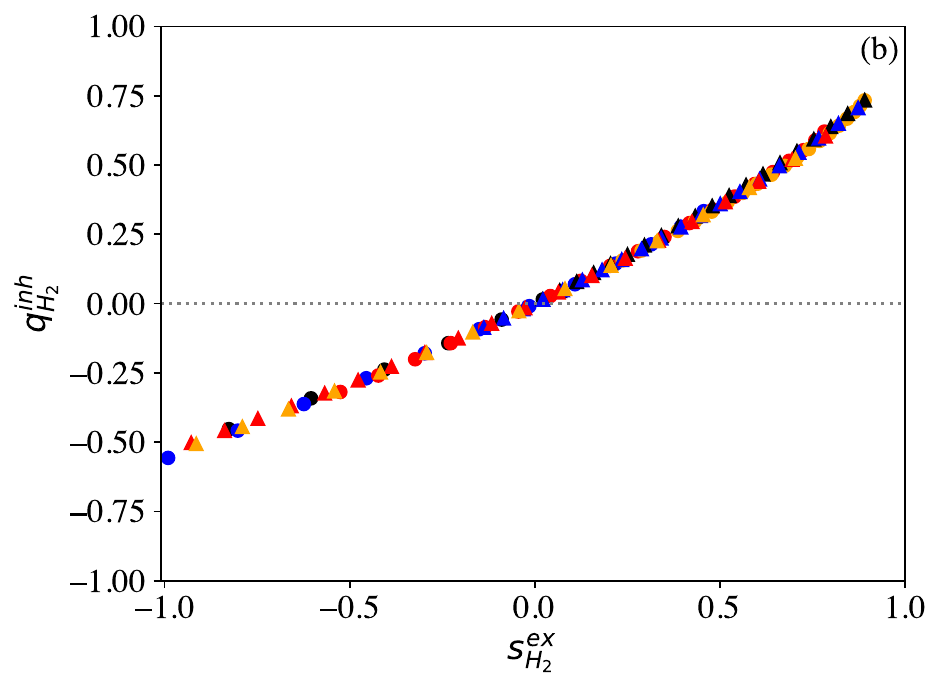}
%plot with empirical curves based on asymptotic solution 
%\includegraphics[width=0.92\columnwidth]{Fig2b_v1.pdf}
%plot of the effectiveness factor empirical curves based on asymptotic solution 
\caption{(a) Hydrogen flux density at the pellet surface for a given
local degree of hydrogenation $c_{+}^{\ex}$ (see legend) 
and varying hydrogen supersaturation at the surface of the
pellet $c_{H_2}^{\ex}$. (b) Hydrogen flux density at the pellet surface for $c_{+}^{\ex}\in [0.05;0.9]$, $c_{H_2}^{\ex}=1,2,5,12$ (black, blue, red, yellow triangles) and for $c_{H_2}^{\ex}\in [1;20]$ $c_{H_2}^{\ex}=0.25,0.5,0.8,0.9$ (black, blue, red, yellow circles), collapsing upon a single curve. In both cases $\Da_+=10$. The other parameters are defined in Eq.~\eqref{eq:default_pars}.}
\label{fig:qh2}
\end{figure}
\begin{figure}[h]
\centering
\includegraphics[width=0.98\columnwidth]{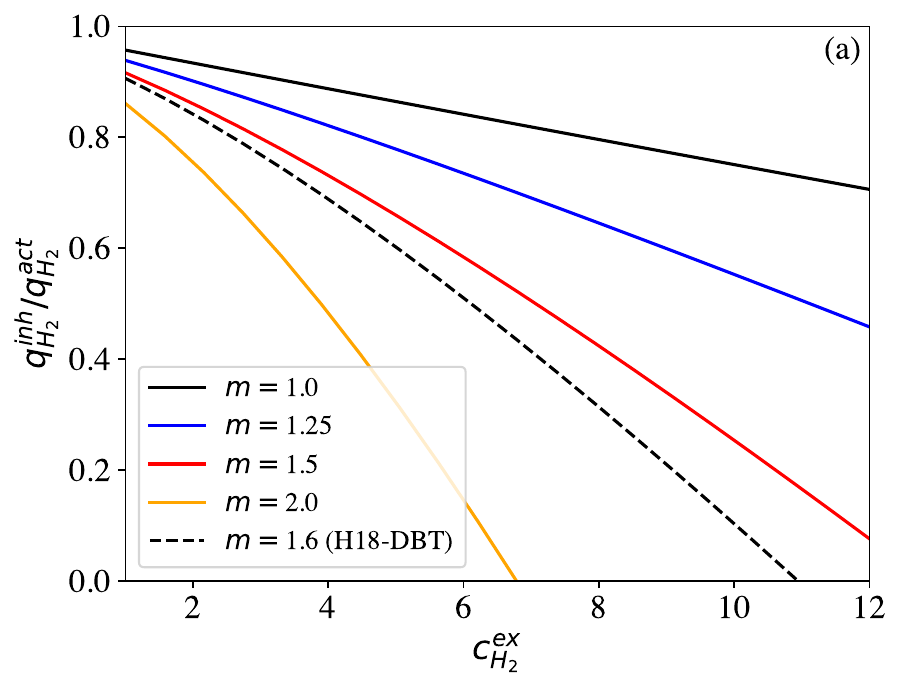}
\includegraphics[width=0.98\columnwidth]{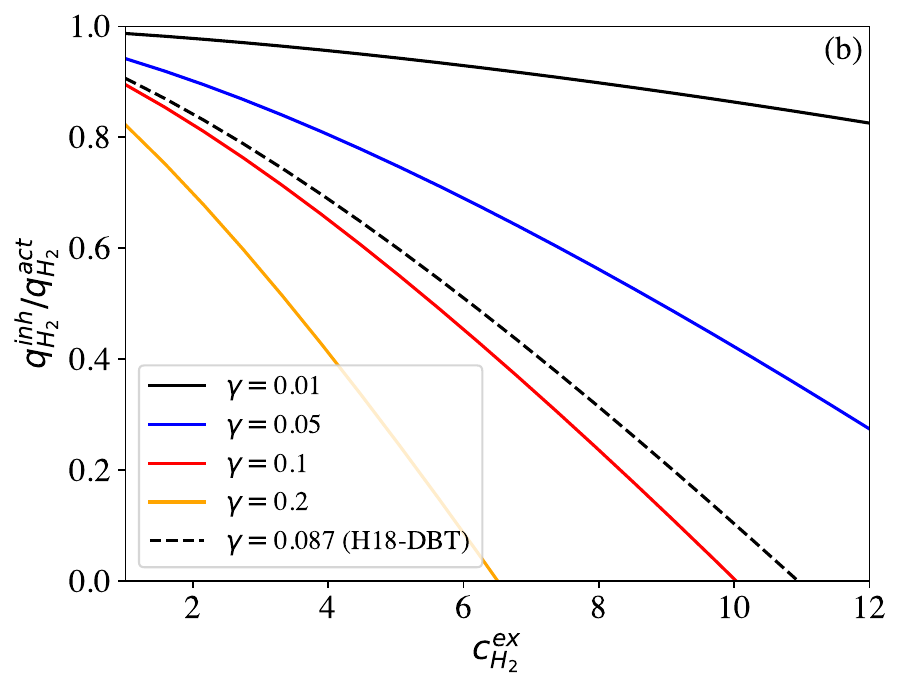}
\caption{Ratio of hydrogen fluxes in the inhibited and active state, depending on the concentration of hydrogen outside of the pellet (a) for a fixed $\gamma=0.087$ and different values of the effective reaction order $m$ and (b) for a fixed $m=1.6$ and different values of $\gamma$. Dashed curves correspond to parameters defined by Eq.\eqref{eq:default_pars} (H18-DBT at $573\,\mathrm{K}$). DoH outside of the pellet is $c_{+}^{\ex}=0.8$, $\Da_+=10$, $\epsilon$ and $h$ are defined in Eq.~\eqref{eq:Da_def}.}
\label{fig:q_ratio}
\end{figure}

Accordingly, we use Eq.~\eqref{eq:qinh_nondim} to calculate the flux of hydrogen $q^{\inh}_{H_2}$ as a function of  hydrogen concentration, $c_{H_2}^{\ex}$, and the local degree of hydrogenation, $c_{+}^{\ex}$, imposed at the surface of the pellet. Fig.~\ref{fig:qh2}(a) shows the non-dimensional hydrogen flux, $q^{\inh}_{H_2}$ as a function of $c_{H_2}^{\ex}$ for some values of $c_+^{\ex}$ and $\Da_+=10$.  As the hydrogen concentration at the pellet surface increases, hydrogen production decreases due to the back reaction. For each degree of hydrogenation, there exists a critical supersaturation level at which hydrogen production stops. Beyond this point, the system switches from dehydrogenation to hydrogenation: the hydrogen flux changes sign. Interestingly, as shown in Fig.~\ref{fig:qh2}(b), this data (as well as additional data points for $c_{H_2}^{\ex}<12$ and $c_{+}^{\ex}<0.9$) collapse onto a master curve when $q_{H_2}$ is plotted as a function of the reaction term, calculated at the pellet surface $s_{H_2}^{\ex}=s_{H_2}(c_{H_2}^{\ex},c_{+}^{\ex})$. This means that productivity of the pellet in this range of parameters is fully defined by the source term at its surface and can be expressed in terms of the effectiveness factor, as defined in  \cite{kao1968effectiveness,rios2023revisiting}.

Let us now calculate the ratio of the fluxes in the active and inhibited state as a function of $c_{H_2}^{\ex}$ at fixed $c_{+}^{\ex}=0.8$. We can see in Fig.~\ref{fig:q_ratio} that at small $c_{H_2}^{\ex}$ the production of hydrogen in the inhibited state is about $10-15\%$ smaller than that in the active state. However, at higher $c_{H_2}^{\ex}$ the ratio of the fluxes becomes very small and turns to zero at some critical level of supersaturation. This critical level depends on the kinetics of the back reaction (LOHC hydrogenation). In Fig.~\ref{fig:q_ratio}(a), we explore the dependency of inhibition on the effective reaction order $m$, characterizing the sensitivity of the back reaction on the hydrogen concentration. As expected, larger values of $m$ lead to stronger inhibition. The same trend is observed for the equilibrium constant of hydrogenation $\gamma$: at larger $\gamma$, the chemical equilibrium is shifted to more hydrogenated LOHC compounds, leading to inhibition at lower levels of $H_2$ supersaturation. Note that $\gamma$ is very sensitive to temperature: for example, for H18-DBT increasing the temperature from $573 \,\mathrm{K}$ to $583 \,\mathrm{K}$ decreases $\gamma$ from $0.087$ to $0.02$~\cite{durr2021experimental}.

Fig.~\ref{fig:q_ratio} clearly shows that the efficient evacuation of hydrogen is a crucial factor for the productivity of the pellet. While in a flow-through reactor hydrogen is advected away by the flow~\cite{uhrig2024reactivation}, hence leading to a small value of $c_{H_2}^{\ex}$, in a small batch experiment hydrogen transport is purely diffusive, hence leading to much larger values of $c_{H_2}^{\ex}$~\cite{solymosi2022nucleation}. In particular, for  $m=1.6$ and $\gamma=0.087$, corresponding to H18-DBT at $573\,\mathrm{K}$ (see Appendix~\ref{sec:params}, dashed curves in Fig.~\ref{fig:q_ratio}) the reaction rate at $c_{H_2}^{\ex}=1$, $c_{+}^{\ex}=0.8$ is just $10\%$ smaller than in the active state, while at $c_{H_2}^{\ex}>10$, $c_{+}^{\ex}=0.8$ the reaction virtually stops. To estimate the values of $c_{+}^{\ex}$ and $c_{H_2}^{\ex}$ in experimental conditions, we need to couple the reaction--diffusion equations inside the pellet to the transport equations outside the pellet, which depend strongly on the experimental setup.

\subsection{Hydrogen production and supersaturation in batch experiments}

So far, we have shown that the hydrogen production within the pellets is very sensitive to both $c_+^{\ex}$ and $c_{H_2}^{\ex}$. However, these parameters cannot be directly controlled in real experimental scenarios. For example, in the batch experiment used to study inhibition, the pellets were placed at the bottom of a test tube filled with hydrogen-rich LOHC compound H18-DBT~\cite{solymosi2022nucleation}. In the following, we estimate the values of $c^{\ex}_{H_2}$ and $c_+^{\ex}$ by modeling the transport of both reactants and products outside the pellets.

In order to keep the model as simple as possible, we assume that the concentrations at the outer surface of the pellet are homogeneous and focus on the diffusion in the liquid layer above the pellets (see Fig.~\ref{fig:sketch}(c)). To estimate the concentrations in the pellet layer, we match the flux of hydrogen produced by the pellets to the diffusive flux across the layer of the LOHC mixture of thickness $L$. Concentrations at the liquid/gas interface are fixed to $c_{H_2}=1$ (equilibrium with gas) and $c_{+}=c_+^{\out}$ (bulk value of the degree of hydrogenation of the LOHC mixture). This leads to the following boundary conditions at the surface of the pellets (see Appendix~\ref{sec:small_batch}):
\begin{eqnarray}
\kappa \dfrac{\partial c_{H_2}}{\partial r}+(c_{H_2}-1)&=0,    \label{eq:bc_kappa}\\
\kappa \dfrac{\partial c_{+}}{\partial r}+(c_{+}-c_{+}^{\out})&=0. \label{eq:bc_kappa_cp}
\end{eqnarray}
\begin{figure}[h!]
\centering
 \includegraphics[width=0.98\columnwidth]{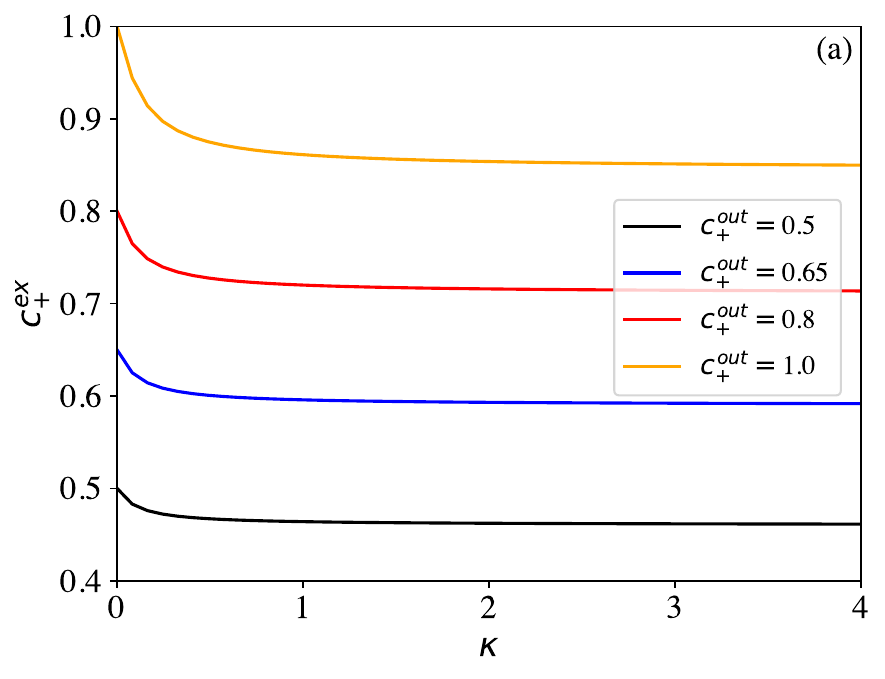}
\includegraphics[width=0.98\columnwidth]{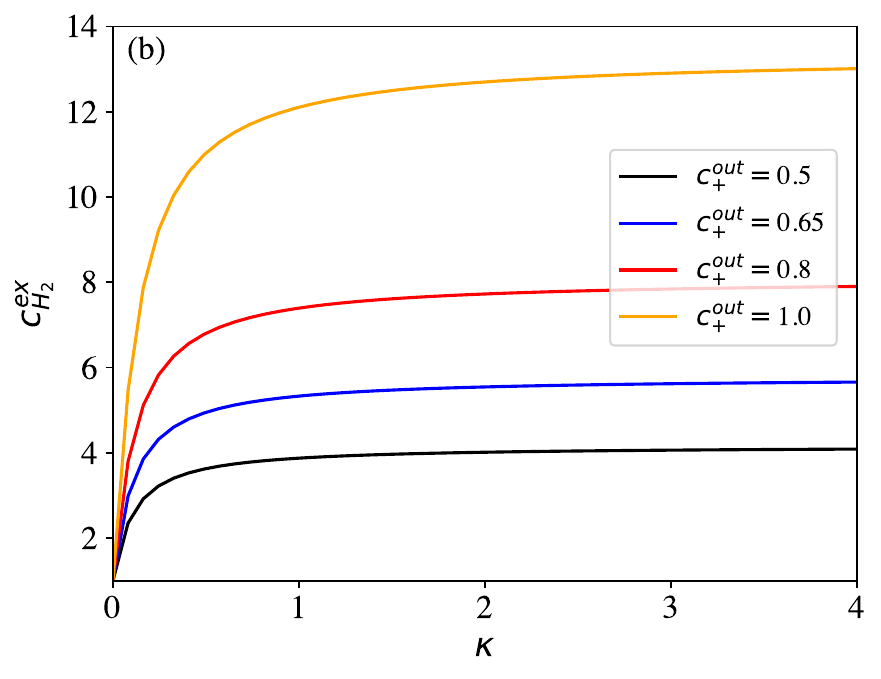}
 \includegraphics[width=0.98\columnwidth]{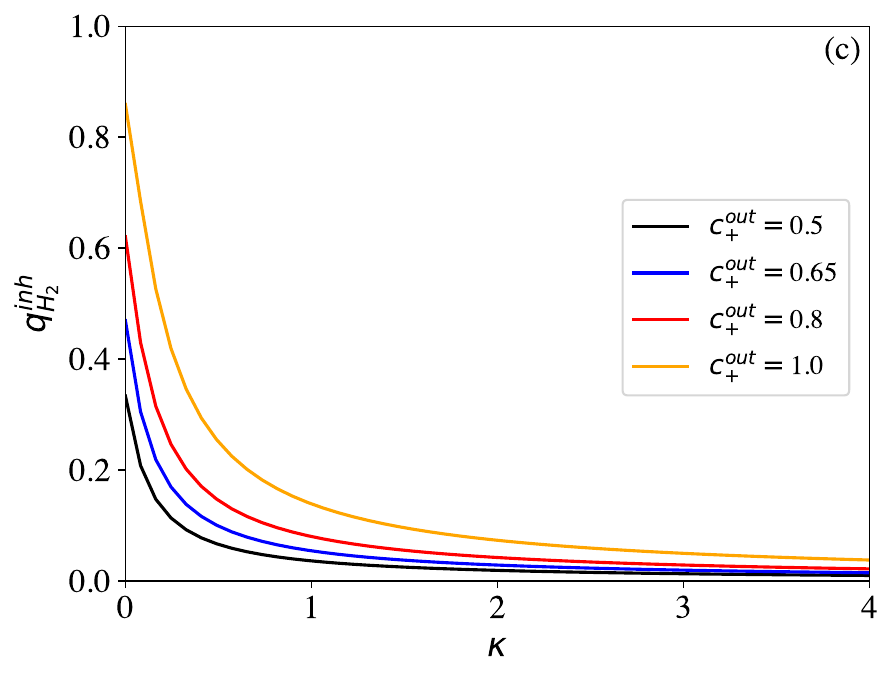}
\caption{Batch setup: (a) Hydrogen supersaturation $c_{H_2}^{\ex}$ and (b) local degree of hydrogenation $c_{+}^{\ex}$ in the pellet layer and (c) hydrogen flux in the inhibited state versus external transport parameter $\kappa$ for different $c_{+}^{\out}$ at $\Da_+=10$. Other parameters are defined in Eq.~\eqref{eq:default_pars}.
}
\label{fig:c_flux_batch}
\end{figure}
Here, we have defined a non-dimensional parameter $\kappa$, responsible for transport outside the pellet as
\begin{equation}\kappa=\psi_{2D}  k_{\eff} \dfrac{L}{R},\end{equation}
where 
\begin{align}
    k_\eff=\dfrac{D_{H_2}^{\eff}}{D_{H_2}}=\dfrac{D_{+}^{\eff}}{D_{+}}
\end{align}
is the ratio of the diffusion coefficients in the porous medium and in the bulk. 
\begin{equation}
    \psi_{2D}=N\cdot \dfrac{4 R^2}{S}
\end{equation}
is the 2D packing density of the pellets, where 
$N$ is the number of pellets in the layer and $S=\pi d^2/4 $ is the cross-section area of the tube with diameter $d$. Note that we assume the correction $k_{\eff}$ to be equal for both $H_2$ and LOHC compounds. In pellets with small pores, these corrections can be different due to the different sizes of the LOHC and $H_2$ molecules~\cite{berezhkovskii2022intrinsic,malgaretti2023closed}, leading to different values of $\kappa$ in Eqs.~\eqref{eq:bc_kappa} and \eqref{eq:bc_kappa_cp}.

Finally, we assume that in the active state, the LOHC mixture in the layer above the pellets is well-mixed due to active bubbling, so that we can use $c_{+}^{\ex}=c_{+}^{\out}$ as a boundary condition for calculations of the active flux.

Solving Eqs.~\eqref{eq:ch2_nd} and \eqref{eq:cp_nd} with the boundary condition given by Eqs.~\eqref{eq:bc_kappa} and \eqref{eq:bc_kappa_cp}, we find $c_{+}^{\ex}$ and $c_{H_2}^{\ex}$ in the pellet layer, and the hydrogen flux in the inhibited state as a function of $\kappa$, $\Da_+$ and $c_{\ex}^{\out}$. 

\begin{figure}[h!]
\centering

\includegraphics[width=0.98\columnwidth]{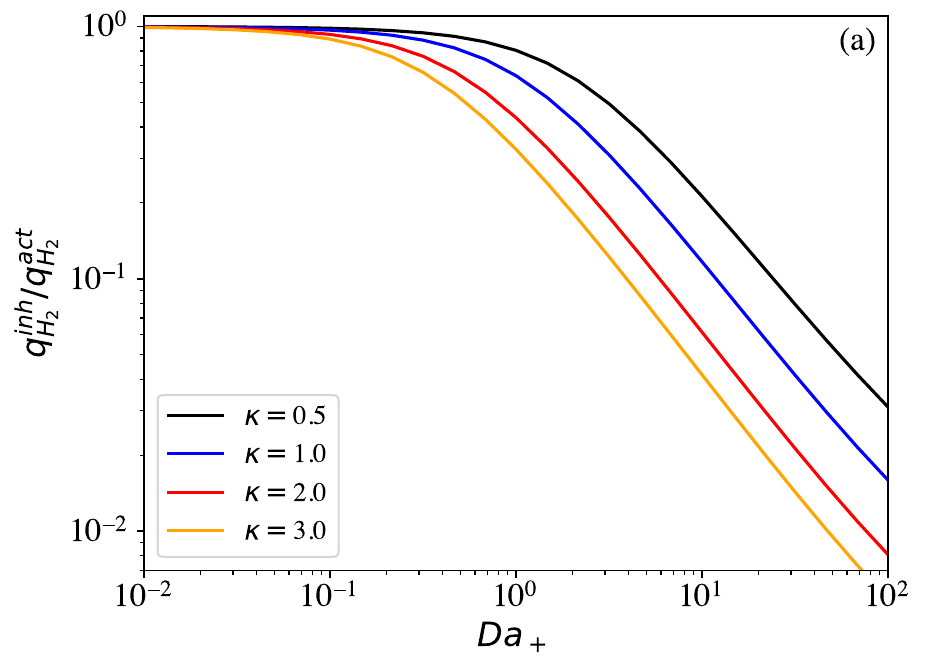}\\
\includegraphics[width=0.98\columnwidth]{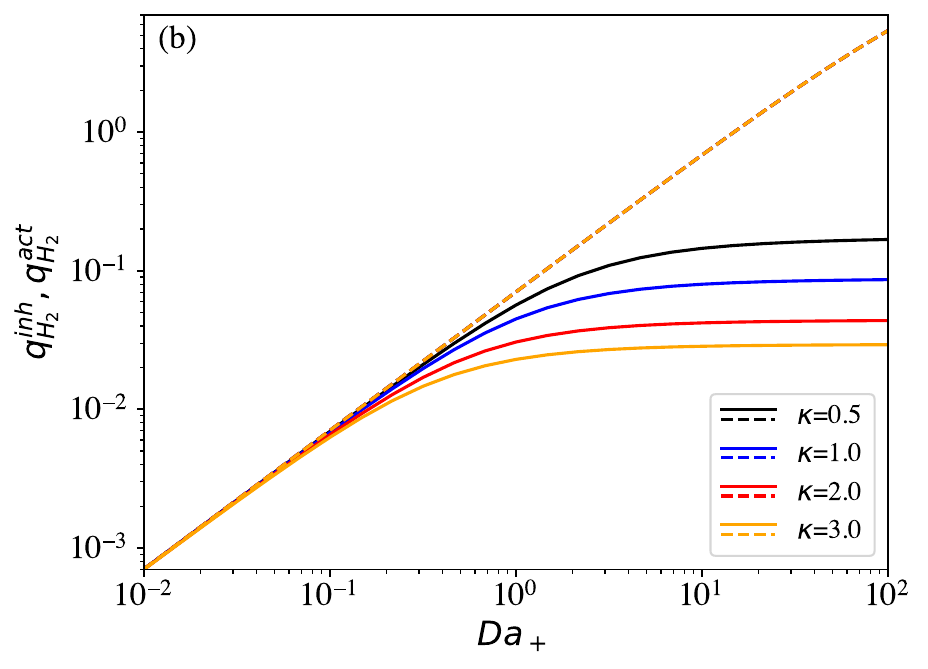}\\
\includegraphics[width=0.98\columnwidth]{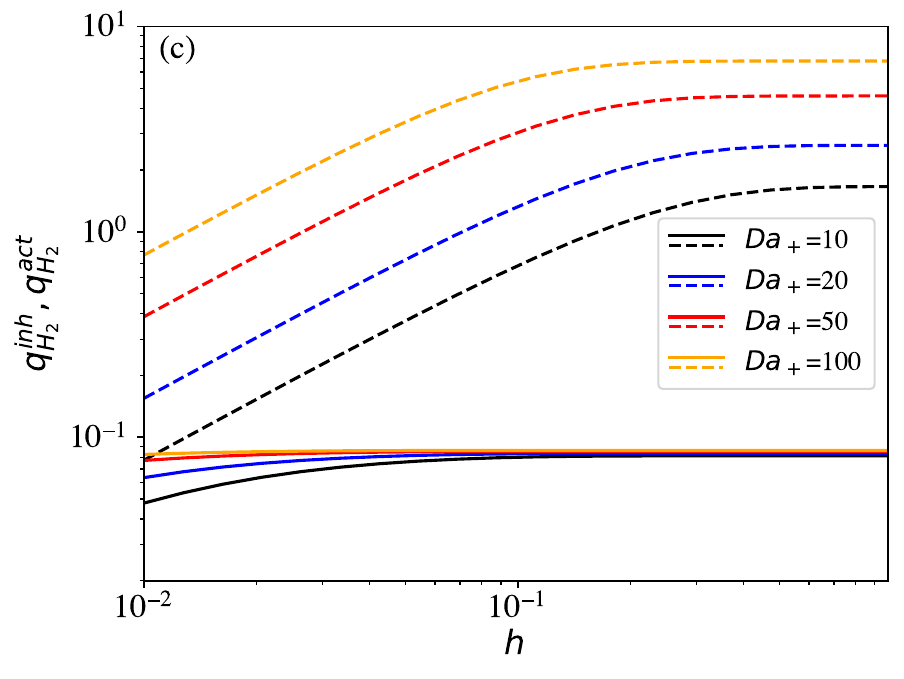}
\caption{Batch setup: (a) Ratio of the fluxes in the active and inhibited states versus $\Da_+$ in a batch setup with $c_+^{\out}=0.8$ and different values of the external transport parameter $\kappa$. (b) Hydrogen flux in the active (dashed, \ta{curves overlap}) and the inhibited (solid) states as a function of $\Da_{+}$ for  $c_+^{\out}=0.8$ and different $\kappa$. (c) Hydrogen flux in the active (dashed) and the inhibited (solid) states as a function of catalytic layer thickness $h$ for $c_{+}^{\out}=0.8$, $\kappa=1$, and varying $\Da_+$. Other parameters are defined in  Eq.~\eqref{eq:default_pars}.
}
\label{fig:qh2_Da_batch}
\end{figure}

Interestingly, as shown in Fig.~\ref{fig:c_flux_batch}(a), the local DoH in the layer, $c_{+}^{\ex}$, does not differ significantly from its bulk value $c_{+}^{\out}$. In contrast, hydrogen supersaturation, $c_{H_2}^{\ex}$, is quite high (see Fig.~\ref{fig:c_flux_batch}(b)) and hydrogen release is hindered by the back reaction. Accordingly, the flux of hydrogen in the inhibited state drops quickly with the increase of $\kappa$ (see Fig.~\ref{fig:c_flux_batch}(c)), showing a high sensitivity of hydrogen production to external transport. 
To estimate the degree of inhibition, we plot in Fig.~\ref{fig:qh2_Da_batch}(a) the ratio of the hydrogen fluxes in the active and inhibited states (inhibition factor) as a function of the Damk\"ohler number $\Da_+$. We consider a fixed $c_+^{\ex}=0.8$ and vary the external transport parameter $\kappa$. We can see that for small $\Da_+$ the fluxes in the active and inhibited states are the same, but their ratio decreases quickly at $\Da_+>1$, following a $\Da_+^{-1}$ scaling. %For experimentally relevant values $\Da_+=10...100$ the inhibition factor does not exceed $0.1$ for $\kappa>0.5$. 
Indeed, we can see in Fig.~\ref{fig:qh2_Da_batch}(b) that the flux of hydrogen in the active state grows linearly with $\Da_+$ (dashed curve), but the flux in the inhibited state (solid curves) is limited. This means that increasing catalyst loading (increasing $\Da_+$) at $\Da_+>10$ does not improve productivity in the inhibited state. 
Interestingly, as shown in Fig.~\ref{fig:qh2_Da_batch}(c), this limiting value is also insensitive to the thickness of the catalytic layer $h$ for $h>0.05$, and it depends only on the external transport parameter $\kappa$. It is important to remark that $\kappa$ depends both on the properties of the pellet (via the correction to the diffusion coefficient $k_\eff$) and on the depth of the liquid layer above the pellets, which can vary from experiment to experiment if not specifically controlled. 

\begin{figure}[h]
\centering
\includegraphics[width=0.98\columnwidth]{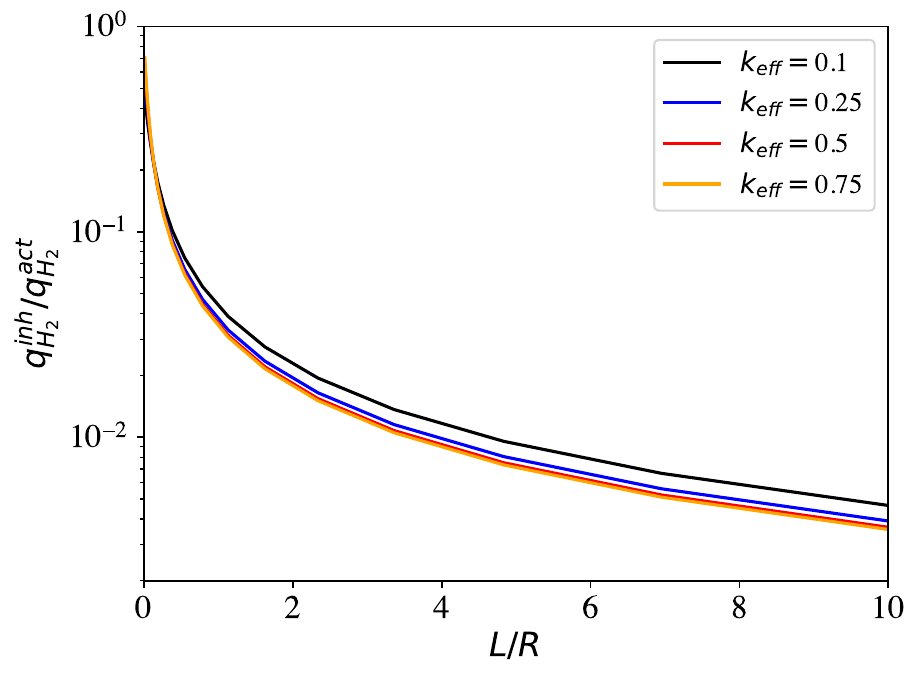}
\caption{Batch setup: ratio of hydrogen flow in inhibited and active state versus the thickness of the liquid layer above the pellets $L/R$ at $c_{+}^{\out}=0.8$, $\Psi_{2D}=0.7$ for different $k_{\eff}$ and fixed $k_{\eff}\Da_+=13$. Other parameters are defined by Eq.~\eqref{eq:default_pars}.}
\label{fig:qh2_batch}
\end{figure}

Let us now compare our predictions with the batch experiment described in Ref.~\cite{solymosi2022nucleation}. For this experiment we have $c_+^{\out}=0.8$, $\psi_{2D}\approx0.7$, $\Da_+\approx 13/k_{\eff}$ and $L/R\approx3.3$ (see Appendix~\ref{sec:small_batch}). While $k_\eff$ for the porous pellets has not been measured, based on the available literature, we expect $k_\eff=0.1..0.75$~\cite{ghanbarian2013tortuosity, sharma1991effective}.  Fig.~\ref{fig:qh2_batch} shows the inhibition factor as a function of the thickness of the LOHC mixture layer $L/R$, for parameters listed above and varying $k_\eff$. We can see that it is not very sensitive to $k_{\eff}$ and its value at $L/R=3$ is between $10^{-2}$ and $10^{-1}$, which is consistent with experimental observations (the flux in the inhibited state is 50 times less than in the active state at DoH=0.8~\cite{solymosi2022nucleation}). %t


\subsection{Inhibited pellet in a flow-through reactor}

\begin{figure}
\centering
\includegraphics[width=0.98\columnwidth]{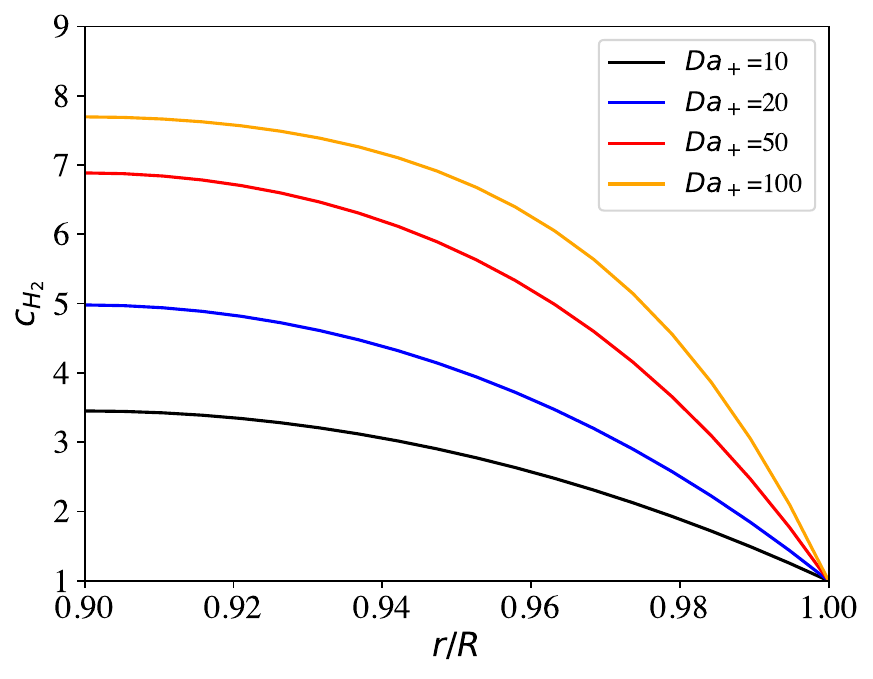}
\caption{Flow-through setup: hydrogen supersaturation in the active layer of the pellet at different values of $\Da_+$ at $c_+^{\ex}=0.8$. Other parameters are given in Eq.~\eqref{eq:default_pars}.}\label{fig:h2_inside}
\end{figure}

As we discussed so far, the dynamics inside the pellet are very sensitive to the supersaturation of hydrogen, $c^{\ex}_{H_2}$, on its surface. However, in batch experiments, this magnitude cannot be directly controlled, and it depends on both the dynamics inside and outside the pellets. This makes the analysis of the dependence of the performance of the pellets on their inner structure complicated since changing the properties of the pellets, such as their porosity (and hence $k$) or the thickness of the catalytic layer, $h$, will also affect the value of $c^{\ex}_{H_2}$. However, contrary to  batch experiments, in flow-through reactors the first layers of pellets are exposed to ``fresh'' LOHC mixture, which does not have any excess dissolved hydrogen, i.e., $c^{\ex}_{H_2}=1$ (see Fig.~\ref{fig:sketch}(d)). 
%Therefore, this a very interesting setup to characterize the dependence of the hydrogen production on the inner properties of the pellets and hence on the dynamics of the hydrogen inside the pellets. 
%\ta{As we discussed so far, the dynamics inside the pellet is very sensitive to the supersaturation of hydrogen, $c^{\ex}_{H_2}$, on its surface. In a batch experiment hydrogen transport in stagnant LOHC outside of the pellet was purely diffusive, which led to build up of high concentrations of hydrogen in the pellet layer. This led to strong inhibition of dehydrogenation reaction, but also to weak sensitivity of the hydrogen flux to catalyst distribution inside the pellet and its inner properties.  However, contrary to  batch experiment, in flow-through reactors the first layers of pellets are exposed to 'fresh' LOHC, which does not have any excess dissolved hydrogen, i.e. $c^{\ex}_{H_2}=1$. 
In this case, the performance of the pellet is limited by hydrogen accumulation inside the pellet itself. Therefore, this is a very interesting setup to characterize the dependence of the hydrogen production on the inner properties of the pellets. Indeed, Fig.~\ref{fig:h2_inside} shows that the density of $H_2$ inside the pellet is sensitive to $\Da_+$, hence raising the issue of the dependence of the performance of the pellet on the inner distribution of hydrogen.

\begin{figure}[h!]
\centering
\includegraphics[width=0.98\columnwidth]{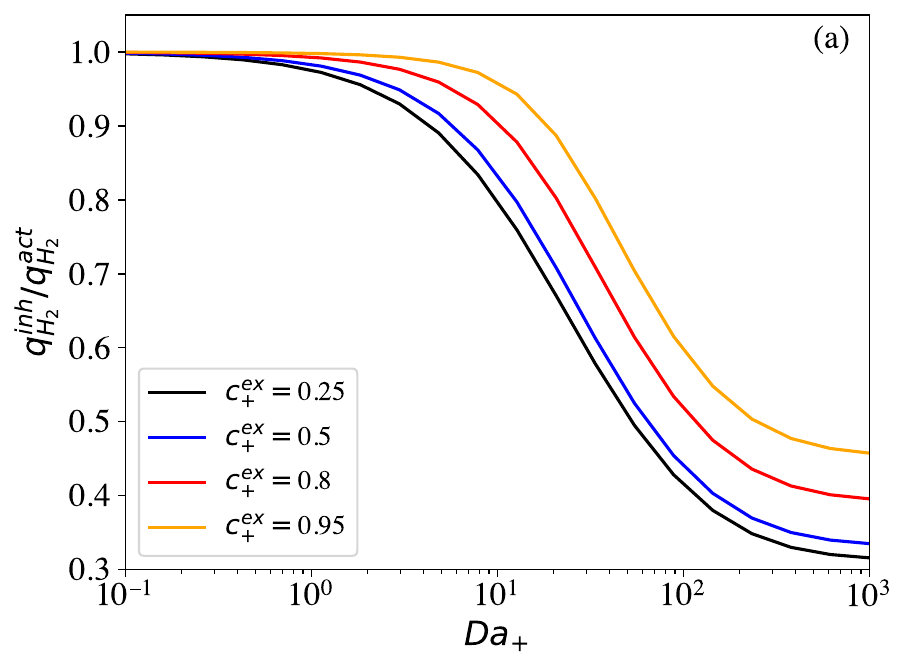}
\includegraphics[width=0.98\columnwidth]{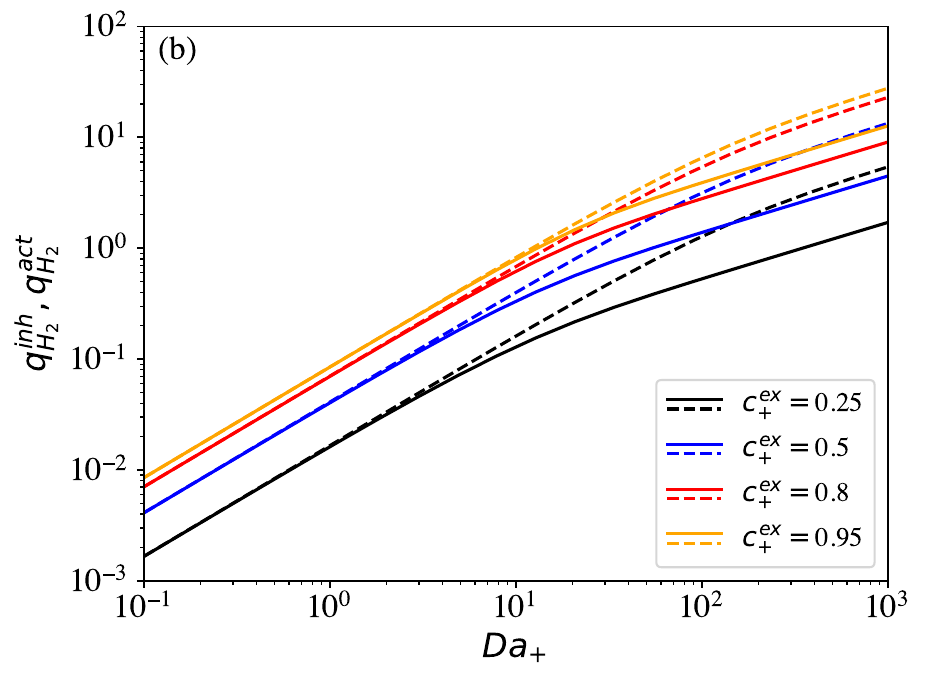}
\includegraphics[width=0.98\columnwidth]{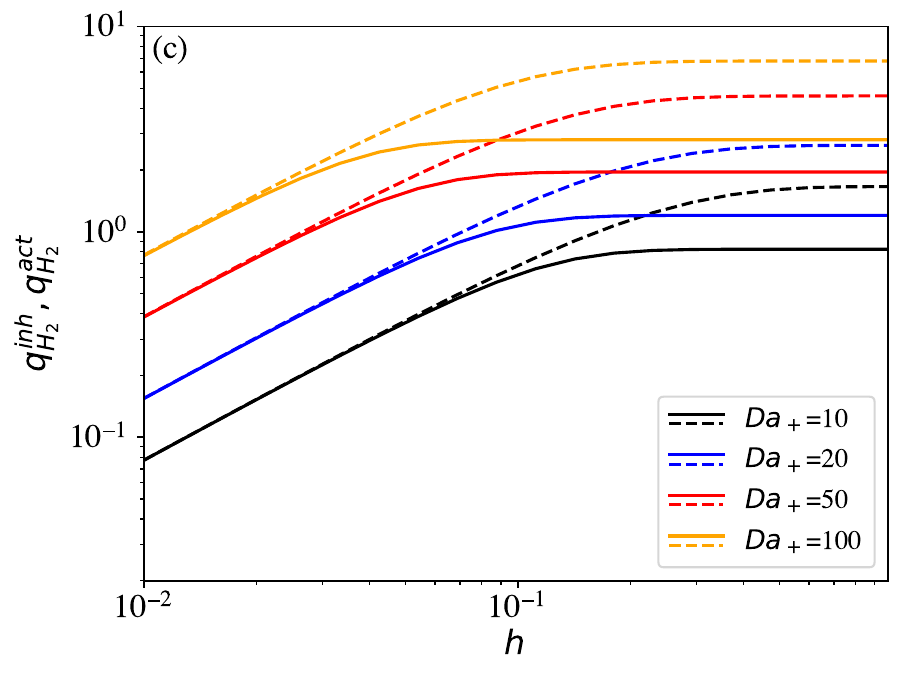}
\caption{Flow-through setup: (a) Ratio of the fluxes in active and inhibited state versus $\Da_+$ for fixed $c_{H_2}^{\ex}=1$ and varying $c_+^{\ex}$. (b) Hydrogen flux in the active (dashed) and the inhibited (solid) states as a function of $\Da_{+}$ for fixed $c_{H_2}^{\ex}=1$ and varying $c_+^{\ex}$. (c) Hydrogen flux in the active (dashed) and the inhibited (solid) states as a function of catalytic layer thickness $h$ for fixed $c_{H_2}^{\ex}=1$, $c_+^{\ex}=0.8$ and varying $\Da_+$. Other parameters are provided in Eq.~\eqref{eq:default_pars}.}
\label{fig:qh2_DaH2} 
\end{figure}

At first, we estimate the degree of inhibition in the flow-through scenario with $c_{H_2}^{\ex}=1$. We plot in Fig.~\ref{fig:qh2_DaH2}(a) the ratio of hydrogen fluxes in the inhibited and active states (inhibition factor) as a function of $\Da_+$ for different $c_+^{\ex}$. It can be seen that the inhibition is stronger for less hydrogenated LOHC mixtures. This is in agreement with experimental observations~\cite{uhrig2024reactivation} and is due to the dependence of the hydrogenation rate on the DoH in Eq.~\eqref{eq:kinetics}: partially dehydrogenated LOHC mixture has more available ``sites'' for hydrogen to attach. We can also see that at small $\Da_+$ the productivity in active and inhibited states is almost the same, while at large $\Da_+$ the ratio of the fluxes decreases. However, the reduction is not as drastic as in the batch case (compare the red curve in Fig.~\ref{fig:qh2_DaH2}(a) and Fig.~\ref{fig:qh2_Da_batch}(a)). 

Indeed, we can see in Fig.~\ref{fig:qh2_DaH2}(b) that, unlike in the batch scenario, the flux in the inhibited state (solid curve) does not saturate at large $\Da_+$, but keeps growing at a sub-linear rate. The flux in the active state (dashed curves) behaves in a similar manner, but the transition to the sub-linear regime occurs at larger $\Da_+$, so that eventually the ratio of the fluxes becomes constant. 

This behavior can be captured by a simple model that takes into account the hydrogen distribution in the pellet. This model can be derived and solved exactly for the case $m=1$ (hydrogenation as a 1st order reaction with respect to hydrogen) and in the limit $\epsilon\ll1$, $h\ll1$ (see Appendix~\ref{sec:asympt}). In this case, an analytical solution for the hydrogen flux can be obtained as
\begin{equation}
    q_{H_2}=-s_{H_2}^{\ex}\lambda^{-1} \tanh(h/\lambda),\label{eq:scaling}
\end{equation}
where
\begin{equation}
    s_{H_2}^{\ex}=c_{H_2}^{\ex}-c_{H_2}^{\eq}(c_{+}^{\ex}),\;  c_{H_2}^{\eq}=\dfrac{c_+^{\ex}}{\gamma(1-c_+^{\ex})}
\end{equation} 
and
\begin{equation}
    \lambda=[\gamma (1-c_+^{\ex})\epsilon^{-1} \Da_{+}]^{-1/2}
\end{equation}
is a non-dimensional length at which chemical equilibrium is attained. The transition from the linear to the sub-linear regime occurs at $\lambda\sim h$, and the transition point depends both on $c_+^{\ex}$ and $h$. Increasing $c_{+}^{\ex}$ shifts the transition to the sub-linear regime (and hence inhibition) to higher values of $\Da_+$, while increasing $h$ shifts the transition to smaller values of $\Da_+$.
%Fig.~\ref{fig:qh2_DaH2}(a) shows the ratio of the fluxes in active and inhibited state depending on $\Da_+$ for different degree of hydrogenation of incoming LOHC (with parameters defined by Eq.~\eqref{eq:default_pars}). \\ 

While this analytical solution is not directly applicable for $m\neq 1$, the model correctly reproduces the scaling of the inhibited flux in the Damk\"ohler number $\Da_+$: it is linear at small $\Da_+$ and proportional to $\sqrt{\Da_+}$ at large $\Da_+$. It also predicts the direction in which the transition point shifts with the change of $c_+^{\ex}$. Indeed, we can see in Fig.~\ref{fig:qh2_DaH2}(b) that the inhibition begins at a smaller $\Da_+$ for LOHC mixtures with a lower degree of hydrogenation $c_{+}^{\ex}$. 

This model also has important implications for the choice of the optimal thickness of the catalytic layer. Eq.~\eqref{eq:scaling} suggests that at $h\gg\lambda$ the flux becomes insensitive to $h$: increasing the thickness of the catalytic layer does not improve the productivity. In Fig.~\ref{fig:qh2_DaH2}(c), we plot the dependence of the hydrogen flux on the thickness of the catalytic layer $h$ for different values of $\Da_{+}$. We can see that, while at small values of $h$ the flux grows linearly, the curves indeed saturate after some critical thickness. This thickness decreases with the growth of $\Da_+$. Pellets with a thin catalytic layer show almost the same productivity in the inhibited state (solid curves in Fig.~\ref{fig:qh2_DaH2}(c)) and in the active one (dashed curves in Fig.~\ref{fig:qh2_DaH2}(c)).

%In particular, for $c_+^{\ex}=0.8$ and $\Da_+=10\ldots100$, inhibition factor in the flow-though scenario is around $0.6-0.9$, while in the batch case it was below $0.1$ 
We can see that the productivity of inhibited catalytic pellets in ideal flow-through conditions %($c_{H_2}^{\ex}=1$) 
can be comparable to that in the active (bubbling) state. As shown in Fig.~\ref{fig:qh2_Da_batch}(c), unlike the batch scenario in flow-through situations, the productivity of the pellet is sensitive to the catalyst distribution in the pellet and to the inner properties of the pellet encoded in the Damk\"ohler number $\Da_+$. The inhibition of hydrogen production is stronger for pellets with a thick catalytic layer and in LOHC mixtures with a lower degree of hydrogenation $c_{+}^{\ex}$.  For $\Da_+=10\ldots50$, typical for dehydrogenation of H18-DBT, we expect the flux in the inhibited state to be $70-90\%$ of the active one (see Fig.~\ref{fig:qh2_DaH2}(a)), in line with what was observed in flow-through experiments~\cite{uhrig2024reactivation}.%, although our model with $c_{H_2}^{ex}=1$ is only applicable to the first few layers of pellets. %Further downstream produced hydrogen will accumulate in the liquid and $c_{H_2}^{ex}$ will increase.
% ### === KEEP THIS, POSSIBLE EXPLANATION ====
%While hydrogen produced by the first layers of pellets accumulates in the solution and may inhibit the reaction downstream, this process is countered by bubble formation in the channel, which take away the excess hydrogen and maintain $c_{H_2}^{\ex}$ low. Indeed, in the flow-through experiment with inhibited pellets~\cite{uhrig2024reactivation} large hydrogen bubbles have been observed throughout the channel and especially in its final sections.

\ta{\section{Transition to bubbling  state}\label{sec:supersat}}
\begin{figure}
\centering
\includegraphics[width=0.98\columnwidth]{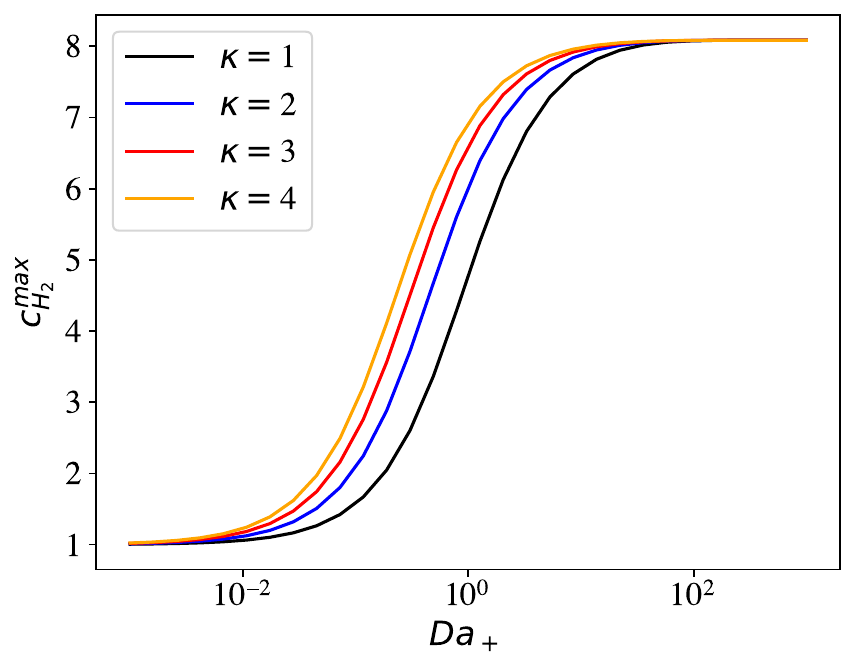}
\caption{\ta{Maximal supersaturation of hydrogen inside the pellets versus $\Da_+$ for a small batch experiment with different values of external transport parameter $\kappa$ and $c_+^{\out}=0.8$.  Other parameters are defined by Eq.~\eqref{eq:default_pars}.}}
\label{fig:ch2_max_batch}
\end{figure}

\begin{figure}
\centering
\includegraphics[width=0.99\columnwidth]{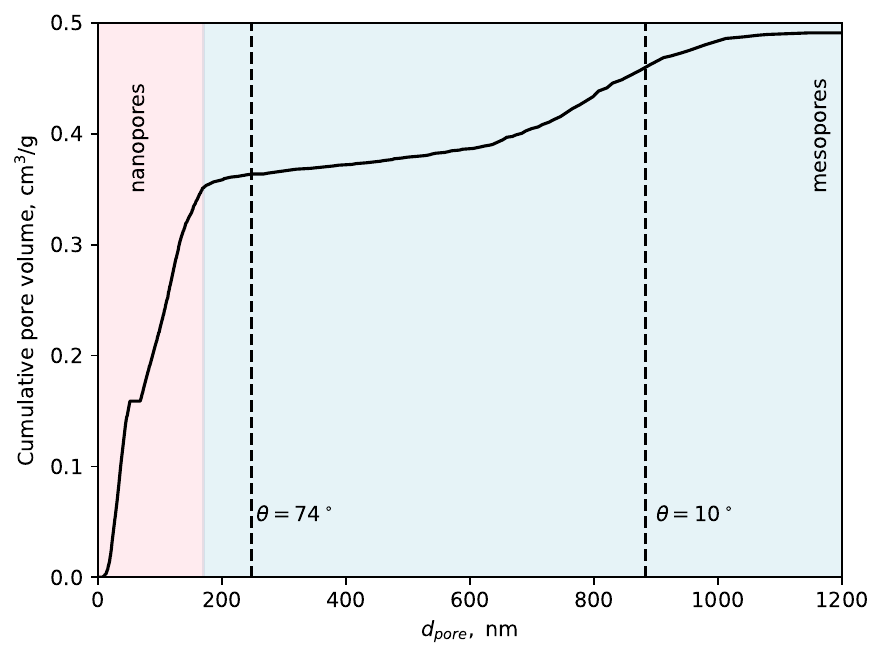}
\caption{\ta{Pore size distribution (cumulative volume) of the catalytic pellet, as from Ref.~\cite{solymosi2022nucleation}. Dashed lines indicate the smallest pore throat that can be passed by growing bubbles at $c_{H_2}=8$, for two different contact angles of the support material.}}
\label{fig:pore_distr}
\end{figure}

\ta{While the porous space of the catalytic pellet provides plenty of sites for heterogenous bubble nucleation~\cite{atchley1989crevice}, growing bubbles need go through a network of pore bottlenecks to form a gas cluster percolating to the surface of the pellet~\cite{LiYortsos1995,li1995theory} (see Fig.~\ref{fig:activation}). Each of these bottlenecks is associated with a capillary entry pressure, which the bubble needs to overcome to pass through~\cite{wang2021capillary,LiYortsos1995}:
\begin{equation}\Delta P=\dfrac{2\sigma}{r_{int}},\quad r_{int}=d_{p}/(2\cos\theta),\end{equation}
where $d_p$ is the pore diameter, $\sigma$ is the surface tension of LOHC and $\theta$ is the receding liquid-side contact angle of the support.
Bubble growth can be sustained only if the surrounding liquid is sufficiently supersaturated with hydrogen to feed the bubble at the required pressure level. Moreover, one can define a critical bottleneck $d_{p}^{cr}$ that has to be passed for isolated bubbles to form a percolating gas cluster, that allows hydrogen to escape from the pellet~\cite{LiYortsos1995}.
The pressure in the gas before passing this critical bottleneck is defined by 
\begin{equation}P_g^{cr}=P_l+\Delta P^{cr},\end{equation}
where $P_l$ is the pressure in liquid and $\Delta P$ is the capillary pressure associated with $d_{p}^{cr}$.}
%Since the gas contains both hydrogen and evaporated LOHC, the total pressure at equilibrium is
%\begin{equation}
%    p_g=p_{H_2}+p_{LOHC}^{LV}(T),
%\end{equation}
%where $p_{H_2}$ is the partial pressure of hydrogen and $p_{LOHC}^{LV}(T)$ is the saturated vapor pressure of the LOHC mixture.
\ta{Assuming that the gas bubble consists only of hydrogen, we obtain then
\begin{equation}
    C_{H_2}^{cr}=H_{PC}P_g=C_{H_2}^{sat}P^{cr}_g/P_{a},
\end{equation}
or, in terms of non-dimensional supersaturation,
\begin{equation}
c_{H_2}^{cr}=\dfrac{P_l}{P_{a}}+\dfrac{4\sigma \cos\theta}{d_{p}^{cr} P_{a}}.
\label{eq:ch2_crit}
\end{equation}
Here, $\sigma$ is the surface tension of LOHC, $d_{p}^{cr}$ is the size of a critical bottleneck that a bubble needs to pass to reach the surface and $P_{a}$ is the atmospheric pressure.}

\ta{Note that the critical supersaturation depends on the size and morphology of the pores via $d_{p}^{cr}$, but also on the wetting properties of the support via the contact angle $\theta$.  Whether this supersaturation can be achieved depends on the balance between production and diffusion of hydrogen and eventually on chemical equilibrium of the reversible reaction. }

%\ta{In the following section we will calculate local supersaturation in the pellet using the reaction-diffusion model and will estimate the fraction of pores accessible to gas from the pore size distribution. We shall see that this fraction varies significantly depending on the wetting properties of the support, leading to inhibited state for 'LOHC-philic' support and spontaneous activation for 'LOHC-phobic' one \cite{solymosi2022nucleation}.}

\iffalse
\ta{Note that hydrogen concentration inside the pellet can not exceed the equilibrium value corresponding to local degree of hydrogenation. Since during dehydrogenation DoH inside the pellet is smaller than at the surface $c_{+}<c_{+}^{\ex}$, we can estimate:
\begin{equation}
    c_{H_2}<c_{H_2}^{\eq}(c_{+}^{\ex})%=\left(\dfrac{c_+^{\ex}}{\gamma(1-c_+^{\ex})}\right)^{1/m}. 
    \label{cH2_eq}
\end{equation}
If this value is smaller than critical supersaturation needed form a percolating gas cluster in the porous space of the pellet, the system will remain in the 'inhibited' state (high hydrogen concentration near the active sites and small dehydrogenation rate) unless triggered by an external stimulus. Note, that both active and inhibited state can exist at the same thermodynamic point (temperature and pressure), because the \textit{local} conditions at catalytic sites are very different due to different mechanisms of hydrogen transport.}
\fi

%\ta{\subsection{Activation conditions}}\label{sec:supersat}

\ta{To estimate whether capillary trapping of bubbles is a plausible mechanism of inhibition,  let us estimate maximal supersaturation $c^{max}_{H_2}$ for the small batch experiment described in~\cite{solymosi2022nucleation}, and compare the corresponding capillary bottleneck to size of the mesopores in the pellet material, provided in~\cite{solymosi2022nucleation}. Fig.~\ref{fig:ch2_max_batch} shows the dependency of maximal supersaturation on $\Da_+$ for different values of external transpot coefficient $\kappa$. Upon increasing the catalyst loading, i.e. $\Da_+$, hydrogen concentration in the pellet increases until the maximal supersaturation is reached.  We can see that in our system, regardless of $\Da_+$ and $\kappa$, supersaturation can not exceed the limiting value $c_{H_2}^{max}\approx 8$. }

\ta{Now let us estimate the smallest pore throat that can be passed by bubbles growing at this supersaturation:
\begin{equation} d_{p}^{min}=\dfrac{4\sigma(T)\cos(\theta)}{(c_{H_2}^{max}-1)P_{a}}\label{eq:dp}
\end{equation}
We can see that it depends strongly on the wetting properties of the support. Indeed, one of the reactivation strategies proposed in~\cite{solymosi2022nucleation} involved chemical modification of the pellets. While for untreated pellets (contact angle $\theta=10^\circ$, $d_p^{min}=880\;nm$) a persistent inhibited state was observed, while perfluoro-modified pellets (contact angle $\theta=74^\circ,\,d_p^{min}=250\; nm$) activated spontaneously. Plotting these values (dashed lines) on top of the pore size distribution of the pellets (see Fig.~\ref{fig:pore_distr}), we can see that for $\theta=10^\circ$ only a small fraction of mesopores can be passed by growing bubbles, while for $\theta=74^\circ$ almost all of mesopore network is accessible.}

\ta{This is consistent with our hypothesis that reversibility of dehydrogenation limits hydrogen concentration in the pellet, leading to capillary trapping of bubbles. Changing wetting properties of the support facilitates bubble escape by reducing the capillary pressure in pores, while overheating (another strategy for pellet reactivation~\cite{solymosi2022nucleation}) shifts chemical equilibrium to less hydrogenated forms of LOHC, allowing for higher $H_2$ concentrations. In contrast, increasing catalyst loading is not efficient as reactivation strategy, because maximal supersaturation is limited by chemical equilibrium (see Fig.\ref{fig:ch2_max_batch}). }

%Note, that presence of excess dissolved hydrogen in the inhibited state and its influence on productivity is also confirmed by experimental evidence: when a LOHC-phobic PTFE cube was placed near the pellet, bubbles nucleated on the cube surface and hydrogen flux was increased twofold compared to the inhibited state without the cubes~\cite{solymosi2022nucleation}. Indeed, placing a nucleation site near the pellets shortens the diffusion path to the liquid/gas interface, decreases local hydrogen concentration and enhances the reaction. Also, when the pellet is reactivated by a mechanical stimulus, hydrogen production peaks briefly before returning to normal level~\cite{solymosi2022nucleation}. This can be attributed to sudden release of hydrogen stored in oversaturated solution prior to reactivation.

\vspace{20pt}
\section{Conclusion}

We derived a model for the reversible hydrogenation/dehydrogenation and the transport of LOHC compounds and hydrogen in inhibited (non-bubbling) catalyst pellets.
Our analysis shows that hydrogen production in the inhibited state depends strongly on the transport of dissolved hydrogen outside the pellet as well as the back reactions. Thus, in a small batch experiment, \ta{due to the slow diffusive transport, } 
%when transport is dominated by diffusion, 
hydrogen production in the absence of bubbling can be 50-100 times smaller than in the active state, which is in good agreement with experimental observations~\cite{solymosi2022nucleation}. 
%The limiting values of supersaturation estimated for H18-DBT are also consistent with the hypothesis of blockage of bubble growth at bottlenecks of the pore space, in agreement with~\cite{solymosi2022nucleation}. 
In contrast, in a flow-through scenario, \ta{the strong advection of the hydrogen outside the pellet leads to a small concentration of hydrogen close to the pellet, and hence}  %when the hydrogen concentration near the pellet is small, 
the difference between the nucleation-inhibited (non-bubbling) and the active (bubbling) state is much smaller and decreases with the degree of hydrogenation of incoming LOHC mixture, which is also consistent with experimental observations~\cite{uhrig2024reactivation}.

%We have derived a simple model that captures the essence of both the chemical reactions occurring the thin shell filled of catalyst as well as the transport of both the reactants and the products.  By explicitly accounting for the back reaction, we have shown that indeed, the latter is crucial in determining the net hydrogen production of pellets in the "inhibited" state, i.e., pellets that have not yet undergone the bubbling transition. \\
%In particular, our result show that the supersaturation of hydrogen inside the pellet becomes insensitive to the catalyst loading once $\Da\gtrsim 1$ and that in such a regime it is also insensitive to the transport properties of the porous support and catalyst distribution in the pellet (thickness of the active shell).
%Quantitative estimates of supersaturation, obtained using experimentally measured parameters for H18-DBT at $T=573K$ are compatible with the hypothesis of inhibition formulated in \st{Section I} Ref.\cite{solymosi2022nucleation}: inability of gas bubbles to form a percolating cluster due to insufficient supersaturation. \\

\ta{Finally, we formulated the conditions for spontaneous activation of the pellet based on local hydrogen concentration. Bubble growth in porous medium is restricted by capillary pressure \cite{li1995theory}. The level of hydrogen supersaturation inside the pellet defines the maximal pressure that can be created in growing bubbles, and hence the size of the smallest pore bottleneck they can pass. If supersaturation is too small, gas bubbles remain trapped by capillary forces and the pellet can not reactivate spontaneously. We estimated hydrogen supersaturation inside the pellet in realistic experimental conditions, compared the corresponding bottleneck size to typical size of the pellet's mesopores measured in \cite{solymosi2022nucleation} and found it consistent with capillary trapping hypothesis.}

Overall, our model shows that the inhibition of dehydrogenation in the presence of dissolved hydrogen (back hydrogenation reaction) is the key mechanism in pellet deactivation. Indeed, in the absence of bubbles, the transport of hydrogen inside the pellet is dominated by diffusion, \ta{which is slow as compared to the reaction rate and it} leads to the build-up of produced hydrogen inside and around the catalytic pellet. 
%This is confirmed by experimental evidence: bubble nucleation on a PTFE cube placed near an inhibited pellet and overshoot in hydrogen production upon pellet reactivation~\cite{solymosi2022nucleation}. 
A high local concentration of hydrogen promotes the back hydrogenation reaction and thus reduces the overall dehydrogenation rate. \ta{The eventual hydrogen concentration is limited by chemical equilibrium and might be insufficient to ensure bubble growth through the bottlenecks of the porous space, preventing spontaneous activation}.  

This effect can also be relevant for the active (bubbling) state. In this paper, we used a simple model of the active state, assuming that, once the gas percolates to the surface of the pellet, bubbles grow at atmospheric pressure (see Fig.~\ref{fig:activation}). However, in reality the pressure in the bubbles changes during their growth/detachment cycle~\cite{wang2021capillary}, leading to the increase of average hydrogen concentration at catalytic sites. Moreover, the time-averaged bubble pressure will likely depend on the typical size of mesopores of the pellet.  Indeed, it has been recently shown that the productivity of catalytic pellets for LOHC hydrogenation can be significantly enhanced by tuning the structure of the porous support~\cite{auer2025enhancing}. \ta{The reaction-diffusion model developed in this paper can be used for the active state once it is complemented with evaporation terms and coupled with the growth equation for the pressure in the gas cluster.}

Here, we focused our attention on the case of LOHC hydrogenation/dehydrogenation due to its relevance in the energy transition. However, the impact of back reactions and transport of volatile products on the performance of catalytic pellets is quite general, and we expect it to be observable also for other chemical reactions.  \ta{For example,   4-NiP reduction (commonly used as a test reaction for catalytic systems) displays different kinetics in bubbling and non-bubbling regimes~\cite{nizkaia2026effects}.}
\ta{Therefore, the next important  step would be to predict quantitatively the level of supersaturation, at which the transition from non-bubbling to bubbling state occurs. To do so one needs a quantitative model of gas percolation on the mesopore network in presence of a reversible chemical reaction, which is still lacking.}

\section*{Acknowledgments}
We acknowledge funding by the Deutsche Forschungs-
gemeinschaft (DFG, German Research Foundation)—Project
No. 431791331—SFB 1452. Furthermore, we acknowledge the
Helmholtz Association of German Research Centers (HGF) and the
Federal Ministry of Education and Research (BMBF), Germany,
for supporting the Innovation Pool project “Solar H$_2$: Highly Pure
and Compressed”.
\section*{Author Declarations}
\subsection*{Conflict of Interest}

The authors have no conflicts of interest to disclose.

\subsection*{Author Contributions}
T. Nizkaia: Conceptualization; model development; programming; data analysis; writing – original draft.
T. Solymosi: Conceptualization; interpretation of experimental insights.
P. Malgaretti: Data analysis; writing – review and editing.
J. Harting: Supervision; funding acquisition; writing – review and editing.
P. Wasserscheid: Supervision; writing – review and editing.
\section*{Data availability}
The data that supports the findings of this study (Python source code used to generate the data) is openly available at: https://doi.org/10.5281/zenodo.17962259. 
\appendix
\section{Parameter estimation for H18-DBT}\label{sec:params}
We estimate the parameters of the model for the conditions reported in the experiments where inhibition was observed~\cite{uhrig2024reactivation, solymosi2022nucleation}.
The LOHC system used in these experiments is H0-DBT/H18-DBT (dibenzyltoluene/perhydro-dibenzyltoluene), which can bind up to 18 hydrogen atoms (equivalently: $n=9$ $H_2$ molecules) and the working temperature is $T=573 \,\mathrm{K}$~\cite{solymosi2022nucleation}. The binary diffusion coefficient $D_+$ of H0-DBT in H18-DBT at this temperature is available from direct measurements~\cite{heller2016binary}:
\begin{equation}
D_+=3\cdot 10^{-9} \;m^2/s    
\end{equation}
The solubility of $H_2$ in H18-DBT is extrapolated from measurements at lower temperatures using the fit proposed in~\cite{aslam2016measurement}:
\begin{equation}\dfrac{C_{H_2}^{\sat}}{\bar{C}}=4.8\cdot 10^{-3}\;\text{mol}_{H_2}/\text{mol}_{\text{H18-DBT}}\end{equation}
No measurements are available for the diffusion coefficient of $H_2$ in H18-DBT, but we expect it to be a little lower than the value measured for a similar compound with smaller molecules (H12-DMP)~\cite{bioucas2020thermal}:
\begin{equation}D_{H_2}<7\cdot 10^{-8}\;m^2/s\end{equation}
We can therefore estimate the parameter $\epsilon$ defined by Eq.~\eqref{eq:eps} to be
\begin{equation}\epsilon=\dfrac{x_{H_2}}{n}\dfrac{D_{H_2}^\eff}{D_{+}^{\eff}}<0.0125\,.\end{equation}
A typical $\Da_{+}$ number can now be estimated from a small batch experiment as described in Ref.~\cite{solymosi2022nucleation}.
The radius of the pellet in the experiment is $R=1.5$ mm, and the width of the active shell, in which the catalyst is distributed, is estimated from SEM images as $H=0.1R$~\cite{solymosi2022nucleation}. 

The flux of $H_2$ per pellet in the active state was measured to be $q^{\act}_{H_2}=1.5\cdot 10^{-6}\;[mol_{H_2}/s]$ at full hydrogenation $\DoH=1$~\cite{solymosi2022nucleation}. Assuming a uniform distribution of catalyst in the outer shell of the pellet, we get the following estimate for the active state at $c_+\equiv1$, $c_{H_2}\equiv1$:
\begin{equation}
k_{\uparrow}>\dfrac{q^{\act}_{H_2}}{n\bar{C}V_{shell}}=0.017
[s^{-1}]
\end{equation} 
Here, $V_{shell}=\dfrac{4\pi}{3}\left[R^3-(R-H)^3\right]$ is the volume of the active shell.
This yields the following estimate for the Damk\"ohler number $\Da_+$:
\begin{equation}
\Da_+=\dfrac{k_{\uparrow} R^2}{D_+^{\eff}}>13/k_\eff\sim 10..100
\end{equation} 
$k_{\eff}=D_+^{\eff}/D_+$ is the correction to the diffusion coefficient of LOHC compounds in a porous medium and typical values are $k_{\eff}=0.1\dots0.8$~\cite{tartakovsky2019diffusion}.
\iffalse
and
\begin{equation}
\Da_{H_2}=\dfrac{k_{\uparrow} R^2}{D_{H_2}^{\eff}}>0.5/k_\eff\sim 0.5...10.\end{equation}

Rescaling to the thickness of the active layer, we get for the LOHC mixture:
$$\Da_+^H=h^2\Da_+\sim 0.1..1,$$
and for hydrogen:
$$\Da_{H_2}^H=h^2\Da_{H_2}\sim 10..100.$$
We can see therefore, that the process is reaction--limited for LOHC and diffusion-limited for hydrogen.
\fi
\begin{figure}[h]
\centering
\includegraphics[width=0.98\columnwidth]{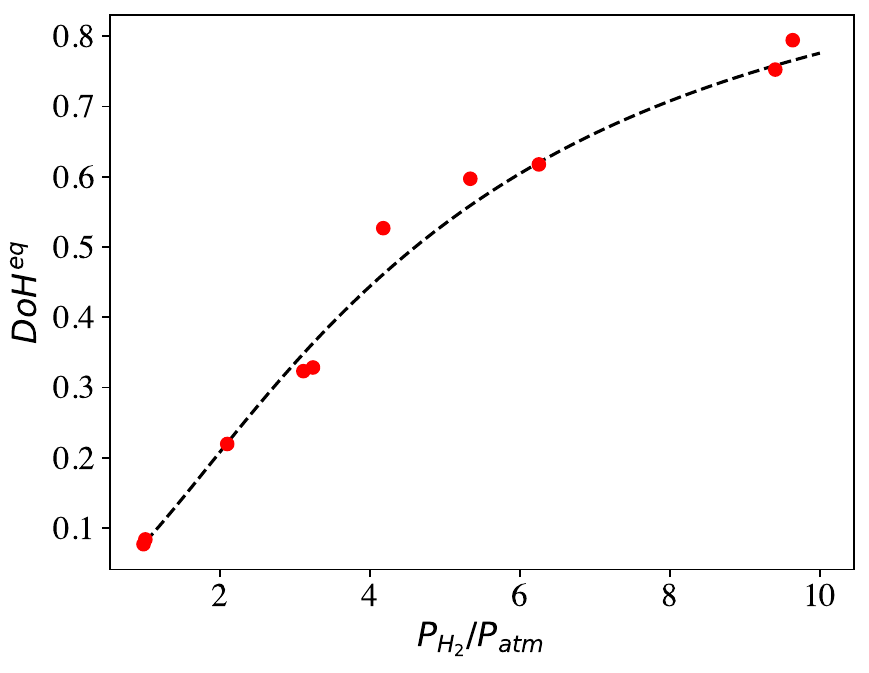}
\caption{Equilibrium $\DoH$ for H18-DBT at $T=573 \,\mathrm{K}$ at different partial pressures of $H_2$: red circles correspond to experimental data from~\cite{durr2021experimental}, the dashed black curve is the approximation by the power law Eq.~\eqref{eq:pow_law} with $\gamma=0.087$, $m=1.6$.}
\label{fig:DoH_eq}
\end{figure}

\begin{figure}[h]
\centering
\includegraphics[width=0.98\columnwidth]{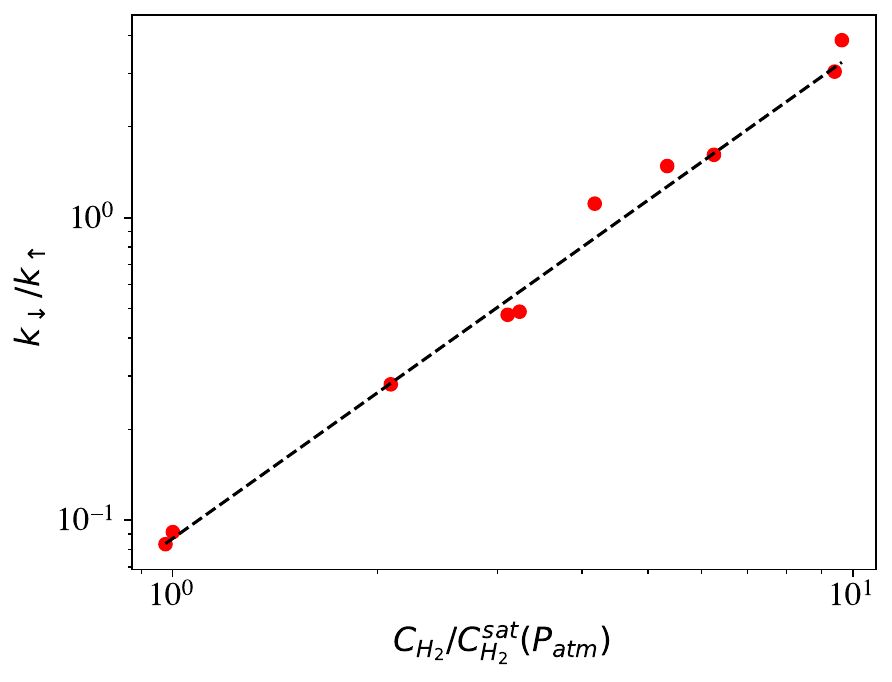}
\caption{Ratio of hydrogenation to dehydrogenation rate for H18-DBT at $T=573 \,\mathrm{K}$: experimental data from Ref.~\cite{durr2021experimental} (red circles) and approximation by a power law Eq.~\eqref{eq:pow_law} (dashed line).}
\label{fig:kup_kdown}
\end{figure}

Finally, the kinetic parameters of the reversible reaction $\gamma$ and $m$ can be estimated from the dependency of the equilibrium degree of hydrogenation $\DoH^{\eq}$ on the partial pressure of hydrogen, available for H18-DBT at $T=573\,\mathrm{K}$ from~\cite{durr2021experimental} (see Fig.~\ref{fig:DoH_eq}). The rate of the back reaction $k_{\downarrow}$ at different partial pressures can be expressed as a function of $\DoH^{\eq}$:
\begin{equation}k_{\downarrow}(P_{H_2})=k_{\uparrow}\cdot\dfrac{\DoH^{\eq}(P_{H_2})}{1-\DoH^{\eq}(P_{H_2})}.
\end{equation}
Since the concentration of dissolved $H_2$ grows linearly with pressure, we can map the pressure to $C_{H_2}$:
\begin{equation}
    \dfrac{C_{H_2}}{C_{H_2}^{\sat}}=\dfrac{P_{H_2}}{P_a}
\end{equation}
We can now fit the dependency $k_{\downarrow}(c_{H_2})$ with a power law:
\begin{equation}
  k_{\downarrow}(C_{H_2}/C_{H_2}^{\sat})/k_{\uparrow}=\gamma\cdot \left(\dfrac{C_{H_2}}{C_{H_2}^{\sat}}\right)^m
  \label{eq:pow_law}\end{equation}
Fig.~\ref{fig:kup_kdown} shows the dependency of $k_{\downarrow}(c_{H_2})/k_{\uparrow}$ for H18-DBT at $T=573\,\mathrm{K}$~\cite{durr2021experimental}, along with the fit using the parameters
\begin{equation}
    \gamma=0.087,\; m=1.6.
\end{equation}
Kinetic analysis of H0-DBT hydrogenation \cite{park2023kinetic} suggests a smaller value for the reaction order ($m=1.2$), but it should be noted that corresponding experiments were performed at lower temperatures ($150^o-170^o$ C) and much higher pressures ($50-70$ bar).

Finally, the surface tension of H18-DBT was taken directly from \cite{solymosi2022nucleation}: 
\begin{equation}
    \sigma=0.123 N/m\text{ at } T=573K
\end{equation}

\section{Boundary conditions and numerical method}\label{sec:small_batch}

Let us derive the boundary conditions for pellets in a small batch experiment: a layer of $N$ pellets immersed in a quiescent LOHC mixture at depth $L$ in a test tube with cross-section area $S$~\cite{solymosi2022nucleation}.  %we can get the following relationship between $c_{H_2}^{\ex}$ and the hydrogen flux $q_{H_2}$.
%on the bottom of a test tube filled with LOHC~\cite{solymosi2022nucleation}.
%For a layer of $N$ pellets 
In the liquid layer above the pellets, we have
\begin{equation}\nabla^2 C_{H_2}=0,\end{equation}
and the condition at the liquid/gas interface is
$$C_{H_2}=C_{H_2}^{\sat}.$$
The solution in the liquid layer is then a linear function,
\begin{equation}C_{H_2}(y)=C_{H_2}^{\sat}+(C_{H_2}^{\ex}-C_{H_2}^{\sat})\cdot K y,\end{equation}
with unknown $K$, which can be found by matching the hydrogen flux produced by the pellets to the diffusive flux across the layer:
\begin{equation}
D_{H_2}\dfrac{C_{H_2}^{\ex}-C_{H_2}^{\sat}}{L}=N\cdot Q_{H_2}/A,\end{equation}
where $A$ is the area of the tube's cross-section.
Here, the flux from one pellet can be calculated as
\begin{equation}Q_{H_2}^{\inh}=4\pi R C_{H_2}^{\sat} D_{H_2}^\eff q_{H_2}.\end{equation}
In non-dimensional variables, we have
\begin{equation}c_{H_2}^{\ex}-1=N \dfrac{4\pi R L}{S}  \dfrac{ D_{H_2}^\eff}{D_{H_2}} q_{H_2}(c_+^{\ex},c_{H_2}^{\ex}),\end{equation}
which can be recast as a boundary condition at the surface of the pellet as
\begin{equation}\kappa \dfrac{\partial c_{H_2}}{\partial r}+(c_{H_2}-1)=0,\end{equation}
with
\begin{equation}\kappa=N \dfrac{4\pi R L}{A}  \dfrac{ D_{H_2}^\eff}{D_{H_2}}.\end{equation}
Introducing a 2D packing density $\psi_{2d}=(\pi R^2 N)/A<1$, we get
\begin{equation}\kappa=4 \psi_{2d} k_{\eff} \dfrac{L}{R},\end{equation}
where $k_\eff=\dfrac{D_{+,H_2}^{\eff}}{D_{+,H_2}}$ is the ratio of the diffusion coefficients in the porous medium and in the bulk (assumed to be the same for LOHC compounds and hydrogen).
Applying the same procedure to $c_+^{\ex}$ with boundary value $c_+^{\ex}=c_+^{\out}$ at the liquid/gas interface, we obtain
\begin{equation}
\kappa \dfrac{\partial c_{+}}{\partial r}+(c_{+}-c_{+}^{\out})=0.
\end{equation}
\begin{figure}[!h]
\centering
\includegraphics[width=0.98\columnwidth]{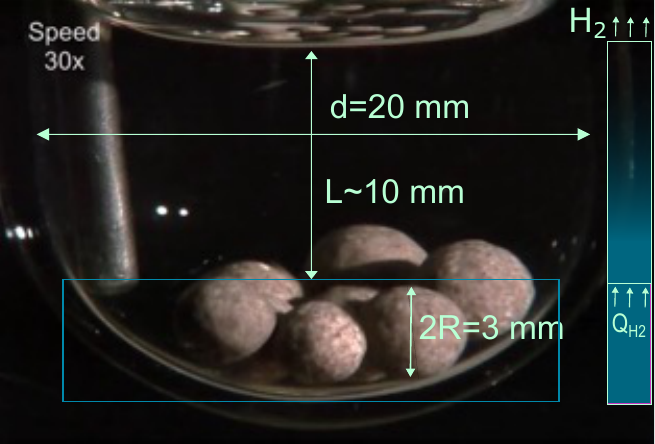}
\caption{A sketch showing experimental setup and defining geometrical parameters from~\cite{solymosi2022nucleation}.}
\label{fig:sketch_exp}
\end{figure}
\iffalse 
For a dense packing of pellets a multi-scale approach can be implemented, following~\cite{qiu2017upscaling}. 

Averaging concentrations $c_{+,H_2}$ over a volume $V$, containing many pellets, we introduce averaged concentrations:
\begin{equation}
    \bar c_{+,H_2}=\int\limits_V c_{+,H_2}(\mathbf x)dV.
\end{equation}
The production of hydrogen in this volume can be obtained by summing up contributions of individual pellets exposed to the same external concentration: $c_{+,H_2}^{\ex}=\bar c_{+,H_2}$. This leads to the following equations for volume-averaged DoH $\bar c_+$ and hydrogen concentration $\bar c_{H_2}$:
\begin{eqnarray}
-\nabla^2\bar c_{H_2}+\lambda \cdot q_{H_2}=0,\label{eq:cH2_out}\\
-\nabla^2\bar c_{+}-\epsilon\cdot\lambda \cdot q_{H_2}=0\label{eq:cp_out}
\end{eqnarray}
where 
\begin{align}
  \lambda=\dfrac{3\psi L^2}{R^2}\dfrac{D_{H_2}^{\eff}}{D_{H_2}^{pack}}, \label{eq:def_lam} 
\end{align}
$\psi$ is packing density (fraction of volume occupied by the pellets), $D_{H_2}^{pack}$ is effective $H_2$ diffusion coefficient in the intra-pellet space, and $L$ is the length scale of the outer problem. \textbf{Sketch for layer, dimensional equations in appendix}
\fi
%To calculate hydrogen supersaturation and local DoH at the surface of the particle  we need to solve the system of reaction--diffusion equations Eqs. (\ref{eq:cp_nd},~\ref{eq:ch2_nd}) with boundary conditions Eq. (\ref{eq:bc1}) or (\ref{eq:bc_kappa}). 

The geometrical parameters of the batch experiment are extracted from the photo in Fig.~\ref{fig:sketch_exp}. The number of pellets in the layer is $N=8$, the depth of the liquid LOHC layer is $L/R\approx3.3$, and the diameter of the tube is $d/R\approx6.5$, so that $\psi_{2d}\approx0.7$. 

\section{Asymptotic estimates.}\label{sec:asympt} 
In catalyst pellets used for LOHC hydrogenation/dehydrogenation, the catalyst is typically distributed in the outer shell of the pellet with a typical thickness $h=H/R\sim 0.1$. In this case, it makes sense to introduce a variable \begin{equation}
    z=r-(1-h)
\end{equation} 
and to solve the equations only in the active layer. No flux boundary conditions are then imposed at the inner boundary of the active zone $z=0$.

%\begin{align}
%  \dfrac{\partial c_+}{\partial z}\Big|_{z=0}=0, \quad \dfrac{\partial c_{H_2}}{\partial z}\Big|_{z=0}=0.
%\end{align}

In the limit $h\ll1$, $\epsilon\ll1$ Eqs.~\eqref{eq:cp_nd} and \eqref{eq:ch2_nd} can be decoupled and solved analytically for $m=1$.
When $\epsilon\ll 1$, $c_+(r)$ changes in space much more slowly than $c_{H_2}(r)$, so that we can consider it as homogeneous.
Indeed, consider the equations 
\begin{eqnarray}
\nabla^2 c_+ -  \Da_{+} \left( c_+ - \gamma c_{H_2}(1 - c_+) \right) = 0,\\
\nabla^2 c_{H_2} + \epsilon^{-1}\Da_{+} \left( c_+ - \gamma c_{H_2}(1 - c_+) \right) = 0,
\end{eqnarray}
with boundary conditions
\begin{equation}
c_+(r=1)=c_+^{\ex},\;c_{H_2}(r=1)=c_{H_2}^{\ex}.\;\end{equation}
In the limit $\epsilon=0$ we have $c_+\equiv c_+^{\ex}$ and the equation for $c_{H_2}$ can be rewritten as
\begin{equation}
\nabla^2 c_{H_2} + \epsilon^{-1}\Da_{+} \gamma (1-c_+^{\ex})\left(c_{H_2}- \dfrac{c_+^{\ex}}{\gamma (1 - c_+^{\ex})} \right) = 0.
\end{equation}
We define a transformed variable as
\begin{equation}
s_{H_2} = c_{H_2} - c_{H_2}^{\eq},
\end{equation}
with the equilibrium hydrogen concentration for given $c_+^{\ex}$
\begin{equation}
    c_{H_2}^{\eq}=\dfrac{c_+^{\ex}}{\gamma (1 - c_+^{\ex})}
\end{equation}
and introduce a non-dimensional parameter $\lambda$ as
\begin{equation}
    \lambda^{-2}=\gamma (1-c_+^{\ex}) \epsilon^{-1}\Da_{+}.
    \end{equation}
The equation then becomes
\begin{equation}
\nabla^2 s_{H_2} - \lambda^{-2} s_{H_2} = 0.
\end{equation}
In a thin layer $h=H/R\ll1$, this equation can be approximated by a 1D equation for a depth variable $z$ with the boundary condition $\dfrac{d s_{H_2}}{d z}=0$ at $z=0$, which can be solved analytically. We get the following solution for $c_+$, $c_{H_2}$:
\iffalse
\begin{equation}
s_{H_2}(z) = \frac{s_{H_2}^{\ex} \cosh(\sqrt{\Da_{H_2}^*} z)}{ \cosh(\sqrt{\Da_{H_2}^* h})}
\end{equation}\fi

\begin{eqnarray}
c_+(z)\equiv c_+^{\ex},\\
c_{H_2}(z)=c_{H_2}^{\eq}(c_+^{\ex})+\frac{s_{H_2}^{\ex} \cosh(z/\lambda)}{ \cosh( h/\lambda)}
\end{eqnarray}
For the hydrogen flux density we obtain
\begin{equation}
    %q_{H_2}=(c_{H_2}^{\eq}-c_{H_2}^{\ex}) \sqrt{\Da_{H_2}^*} \tanh\left(\sqrt{\Da_{H_2}^*}h\right),     
    q_{H_2}(c_+^{\ex},c_{H_2}^{\ex})=s_{H_2}^{\ex}\lambda^{-1}\tanh\left(h/\lambda\right).\label{eq:q_asympt}
\end{equation}
\iffalse
Here 
\begin{equation}
s_{H_2}^{\ex}=c_{H_2}^{\ex}-c_{H_2}^{\eq}(c_+^{\ex})
\end{equation}
is the deviation of hydrogen concentration at the surface of the pellet from its equilibrium value at given $c_{+}$:
\begin{equation}
    c_{H_2}^{\eq}(c_{+}^{\ex})=\dfrac{c_+^{\ex}}{\gamma(1-c_+^{\ex})}
\end{equation}
taken at $m=1$.
Non-dimensional parameter $\lambda$ characterizes the ratio of hydrogenation rate to diffusion rate at fixed $c^{\ex}_+$:
\begin{eqnarray}
  \lambda^2=[(1-c^{\ex}_+) \gamma\Da_{H_2}]^{-1}
\end{eqnarray}
While this solution has been obtained for $m=1$, we expect a similar scaling for the flux for $1<m<2$. In this case, however, it is necessary to solve Eq.(\ref{eq:ch2_nd}) numerically with $c_+\equiv c_{+}^{\ex}$.
The total hydrogen production can be calculated as:
\begin{multline}
Q_{H_2}^{\inh}=4\pi R C_{H_2}^{\sat}D_{H_2}^{\eff}\cdot q_{H_2}\\  q_{H_2}=-\left.\frac{d c_{H_2}}{dr} \right|_{r=1} = (c_{H_2}^{\eq}-c_{H_2}^{\ex}) \sqrt{\Da_{H_2}^*} \tanh\left(\sqrt{\Da_{H_2}^*}h\right)
\end{multline}
\fi
 \bibliography{pellets}

@article{KWAK2021114124,
title = {Hydrogen production from homocyclic liquid organic hydrogen carriers (LOHCs): Benchmarking studies and energy-economic analyses},
journal = {Energy Conversion and Management},
volume = {239},
pages = {114124},
year = {2021},
issn = {0196-8904},
doi = {https://doi.org/10.1016/j.enconman.2021.114124},
author = {Yeonsu Kwak and Jaewon Kirk and Seongeun Moon and Taeyoon Ohm and Yu-Jin Lee and Munjeong Jang and La-Hee Park and Chang-il Ahn and Hyangsoo Jeong and Hyuntae Sohn and Suk Woo Nam and Chang Won Yoon and Young Suk Jo and Yongmin Kim}
}

@article{malgaretti2023closed,
  title={Closed formula for transport across constrictions},
  author={Malgaretti, Paolo and Harting, Jens},
  journal={Entropy},
  volume={25},
  number={3},
  pages={470},
  year={2023},
  publisher={MDPI}
}

@article{berezhkovskii2022intrinsic,
  title={Intrinsic diffusion resistance of a membrane channel, mean first-passage times between its ends, and equilibrium unidirectional fluxes},
  author={Berezhkovskii, Alexander M and Bezrukov, Sergey M},
  journal={The Journal of Chemical Physics},
  volume={156},
  number={7},
  year={2022},
  publisher={AIP Publishing}
}

@article{kao1968effectiveness,
  title={Effectiveness factors for reversible reactions},
  author={Kao, HS-P and Satterfield, CN},
  journal={Industrial \& Engineering Chemistry Fundamentals},
  volume={7},
  number={4},
  pages={664--667},
  year={1968},
  publisher={ACS Publications}
}

@article{heublein_hydrogen_2020,
	title = {Hydrogen storage using liquid organic carriers: {Equilibrium} simulation and dehydrogenation reactor design},
	volume = {45},
	issn = {0360-3199},
	shorttitle = {Hydrogen storage using liquid organic carriers},
	doi = {10.1016/j.ijhydene.2020.04.274},
	number = {46},
	urldate = {2024-05-27},
	journal = {International Journal of Hydrogen Energy},
	author = {Heublein, Norbert and Stelzner, Malte and Sattelmayer, Thomas},
	month = sep,
	year = {2020},
	pages = {24902--24916},
}

@article{jo_recent_2022,
	title = {Recent progress in dehydrogenation catalysts for heterocyclic and homocyclic liquid organic hydrogen carriers},
	volume = {39},
	issn = {1975-7220},
	doi = {10.1007/s11814-021-0947-5},
	language = {en},
	number = {1},
	urldate = {2024-05-27},
	journal = {Korean Journal of Chemical Engineering},
	author = {Jo, Yeongin and Oh, Jinho and Kim, Donghyeon and Park, Ji Hoon and Baik, Joon Hyun and Suh, Young-Woong},
	month = jan,
	year = {2022},
	pages = {20--37},
}

@article{sekine_recent_2021,
	title = {Recent {Trends} on the {Dehydrogenation} {Catalysis} of {Liquid} {Organic} {Hydrogen} {Carrier} ({LOHC}): {A} {Review}},
	volume = {64},
	issn = {1572-9028},
	shorttitle = {Recent {Trends} on the {Dehydrogenation} {Catalysis} of {Liquid} {Organic} {Hydrogen} {Carrier} ({LOHC})},
	doi = {10.1007/s11244-021-01452-x},
	language = {en},
	number = {7},
	urldate = {2024-05-27},
	journal = {Topics in Catalysis},
	author = {Sekine, Yasushi and Higo, Takuma},
	month = jul,
	year = {2021},
	pages = {470--480},
}

@article{taube_system_1983,
	title = {A system of hydrogen-powered vehicles with liquid organic hydrides},
	volume = {8},
	issn = {0360-3199},
	doi = {10.1016/0360-3199(83)90067-8},
	number = {3},
	urldate = {2024-05-27},
	journal = {International Journal of Hydrogen Energy},
	author = {Taube, M. and Rippin, D. W. T. and Cresswell, D. L. and Knecht, W.},
	month = jan,
	year = {1983},
	pages = {213--225},
}

@article{newson_seasonal_1998,
	title = {Seasonal storage of hydrogen in stationary systems with liquid organic hydrides},
	volume = {23},
	issn = {0360-3199},
	doi = {10.1016/S0360-3199(97)00134-1},
	number = {10},
	urldate = {2024-05-27},
	journal = {International Journal of Hydrogen Energy},
	author = {Newson, E. and Haueter, Th. and Hottinger, P. and Von Roth, F. and Scherer, G. W. H. and Schucan, Th. H.},
	month = oct,
	year = {1998},
	pages = {905--909},
}

@article{singla_hydrogen_2021,
        title = {Hydrogen fuel and fuel cell technology for cleaner future: a review},
        volume = {28},
        issn = {1614-7499},
        shorttitle = {Hydrogen fuel and fuel cell technology for cleaner future},
        doi = {10.1007/s11356-020-12231-8},
        journal = {Environmental Science and Pollution Research},
        author = {Singla, Manish Kumar and Nijhawan, Parag and Oberoi, Amandeep Singh},
        year = {2021},
        pages = {15607},
}

@article{staffell_role_2019,
        title = {The role of hydrogen and fuel cells in the global energy system},
        volume = {12},
        doi = {10.1039/C8EE01157E},
        journal = {Energy \& Environmental Science},
        author = {Staffell, Iain and Scamman, Daniel and Abad, Anthony Velazquez and Balcombe, Paul and Dodds, Paul E. and Ekins, Paul and Shah, Nilay and Ward, Kate R.},
        year = {2019},
        pages = {463},
}

@article{LiYortsos1995,
  author  = {Li, X. and Yortsos, Y. C.},
  title   = {Visualization and Simulation of Bubble Growth in Pore Networks},
  journal = {AIChE Journal},
  volume  = {41},
  number  = {2},
  pages   = {214--222},
  year    = {1995},
  doi     = {10.1002/aic.690410203}
}

@article{atchley1989crevice,
  title={The crevice model of bubble nucleation},
  author={Atchley, Anthony A and Prosperetti, Andrea},
  journal={The Journal of the Acoustical Society of America},
  volume={86},
  number={3},
  pages={1065--1084},
  year={1989},
  publisher={Acoustical Society of America}
}

@article{nizkaia2026effects,
  title={Effects of hydrogen transport on the kinetic regimes of 4-nitrophenol reduction by sodium borohydride},
  author={Nizkaia, Tatiana and Groppe, Philipp and M{\"u}ller, Valentin and Harting, Jens and Wintzheimer, Susanne and Malgaretti, Paolo},
  journal={Catalysis Science \& Technology},
  year={2026},
  publisher={Royal Society of Chemistry}
}

@article{solymosi2022nucleation,
  title={Nucleation as a rate-determining step in catalytic gas generation reactions from liquid phase systems},
  author={Solymosi, Thomas and Gei{\ss}elbrecht, Michael and Mayer, Sophie and Auer, Michael and Leicht, Peter and Terlinden, Markus and Malgaretti, Paolo and B{\"o}smann, Andreas and Preuster, Patrick and Harting, Jens and others},
  journal={Science Advances},
  volume={8},
  number={46},
  pages={eade3262},
  year={2022},
  publisher={American Association for the Advancement of Science}
}

@article{durr2021experimental,
  title={Experimental determination of the hydrogenation/dehydrogenation-Equilibrium of the LOHC system H0/H18-dibenzyltoluene},
  author={D{\"u}rr, S and Zilm, S and Gei{\ss}elbrecht, M and M{\"u}ller, Karsten and Preuster, P and B{\"o}smann, A and Wasserscheid, P},
  journal={International Journal of Hydrogen Energy},
  volume={46},
  number={64},
  pages={32583--32594},
  year={2021},
  publisher={Elsevier}
}

@article{aslam2016measurement,
  title={Measurement of hydrogen solubility in potential liquid organic hydrogen carriers},
  author={Aslam, Rabya and M{\"u}ller, Karsten and M{\"u}ller, Michael and Koch, Marcus and Wasserscheid, Peter and Arlt, Wolfgang},
  journal={Journal of chemical \& engineering data},
  volume={61},
  number={1},
  pages={643--649},
  year={2016},
  publisher={ACS Publications}
}

@article{aslam2018thermophysical,
  title={Thermophysical studies of dibenzyltoluene and its partially and fully hydrogenated derivatives},
  author={Aslam, Rabya and Khan, Muhammad Hashim and Ishaq, Muhammad and Mu\"{u}ller, Karsten},
  journal={Journal of Chemical \& Engineering Data},
  volume={63},
  number={12},
  pages={4580--4587},
  year={2018},
  publisher={ACS Publications}
}

@article{bioucas2020thermal,
  title={Thermal conductivity of hydrocarbon liquid organic hydrogen carrier systems: measurement and prediction},
  author={Berger Bioucas, Francisco E and Piszko, Maximilian and Kerscher, Manuel and Preuster, Patrick and Rausch, Michael H and Koller, Thomas M and Wasserscheid, Peter and Fr{\"o}ba, Andreas P},
  journal={Journal of Chemical \& Engineering Data},
  volume={65},
  number={10},
  pages={5003--5017},
  year={2020},
  publisher={ACS Publications}
}

@article{heller2016binary,
  title={Binary diffusion coefficients of the liquid organic hydrogen carrier system dibenzyltoluene/perhydrodibenzyltoluene},
  author={Heller, Andreas and Rausch, Michael H and Schulz, Peter S and Wasserscheid, Peter and Fr\"oba, Andreas P},
  journal={Journal of chemical \& engineering data},
  volume={61},
  number={1},
  pages={504--511},
  year={2016},
  publisher={ACS Publications}
}

@article{kerscher2020thermophysical,
  title={Thermophysical properties of diphenylmethane and dicyclohexylmethane as a reference liquid organic hydrogen carrier system from experiments and molecular simulations},
  author={Kerscher, Manuel and Klein, Tobias and Schulz, Peter S and Veroutis, Emmanouil and D{\"u}rr, Stefan and Preuster, Patrick and Koller, Thomas M and Rausch, Michael H and Economou, Ioannis G and Wasserscheid, Peter and others},
  journal={International journal of hydrogen energy},
  volume={45},
  number={53},
  pages={28903--28919},
  year={2020},
  publisher={Elsevier}
}

@article{sharma1991effective,
  title={Effective diffusion coefficients and tortuosity factors for commercial catalysts},
  author={Sharma, Ramesh K and Cresswell, David L and Newson, Esmond J},
  journal={Industrial \& engineering chemistry research},
  volume={30},
  number={7},
  pages={1428--1433},
  year={1991},
  publisher={ACS Publications}
}

@article{ghanbarian2013tortuosity,
  title={Tortuosity in porous media: a critical review},
  author={Ghanbarian, Behzad and Hunt, Allen G and Ewing, Robert P and Sahimi, Muhammad},
  journal={Soil Science Society of America Journal},
  volume={77},
  number={5},
  pages={1461--1477},
  year={2013},
  publisher={Wiley Online Library}
}

@article{tartakovsky2019diffusion,
  title={Diffusion in porous media: phenomena and mechanisms},
  author={Tartakovsky, Daniel M and Dentz, Marco},
  journal={Transport in Porous Media},
  volume={130},
  pages={105--127},
  year={2019},
  publisher={Springer}
}

@article{uhrig2024reactivation,
  title={Reactivation strategies for nucleation-inhibited catalyst beds in continuously operated gas-release reactions from liquids},
  author={Uhrig, Felix and Solymosi, Thomas and Preuster, Patrick and B{\"o}smann, Andreas and Wasserscheid, Peter},
  journal={International Journal of Hydrogen Energy},
  volume={49},
  pages={1528--1535},
  year={2024},
  publisher={Elsevier}
}

@article{salman2022catalysis,
  title={Catalysis in liquid organic hydrogen storage: Recent advances, challenges, and perspectives},
  author={Salman, Muhammad Saad and Rambhujun, Nigel and Pratthana, Chulaluck and Srivastava, Kshitij and Aguey-Zinsou, Kondo-Francois},
  journal={Industrial \& Engineering Chemistry Research},
  volume={61},
  number={18},
  pages={6067--6105},
  year={2022},
  publisher={ACS Publications}
}

@article{rao2020potential,
  title={Potential liquid-organic hydrogen carrier (LOHC) systems: A review on recent progress},
  author={Rao, Purna Chandra and Yoon, Minyoung},
  journal={Energies},
  volume={13},
  number={22},
  pages={6040},
  year={2020},
  publisher={MDPI}
}

@article{li1995theory,
  title={Theory of multiple bubble growth in porous media by solute diffusion},
  author={Li, X and Yortsos, YC},
  journal={Chemical engineering science},
  volume={50},
  number={8},
  pages={1247--1271},
  year={1995},
  publisher={Elsevier}
}

@article{teichmann2011future,
  title={A future energy supply based on Liquid Organic Hydrogen Carriers (LOHC)},
  author={Teichmann, Daniel and Arlt, Wolfgang and Wasserscheid, Peter and Freymann, Raymond},
  journal={Energy \& Environmental Science},
  volume={4},
  number={8},
  pages={2767--2773},
  year={2011},
  publisher={Royal Society of Chemistry}
}

@article{rakic2023liquid,
  title={Liquid organic hydrogen carrier hydrogenation--dehydrogenation: from ab initio catalysis to reaction micro-kinetics modelling},
  author={Raki{\'c}, Emilija and Grilc, Miha and Likozar, Bla{\v{z}}},
  journal={Chemical engineering journal},
  pages={144836},
  year={2023},
  publisher={Elsevier}
}

@article{jorschick2017hydrogen,
  title={Hydrogen storage using a hot pressure swing reactor},
  author={Jorschick, Holger and Preuster, Patrick and D{\"u}rr, Stefan and Seidel, Alexander and M{\"u}ller, Karsten and B{\"o}smann, Andreas and Wasserscheid, P},
  journal={Energy \& environmental science},
  volume={10},
  number={7},
  pages={1652--1659},
  year={2017},
  publisher={Royal Society of Chemistry}
}

@article{shi2023dehydrogenation,
  title={Dehydrogenation of the liquid organic hydrogen carrier perhydrodibenzyltoluene--reaction pathway over Pt/Al 2 O 3},
  author={Shi, Libin and Qi, Suitao and Smith, Kevin J and Alamoudi, Majed and Zhou, Yiming},
  journal={Reaction Chemistry \& Engineering},
  volume={8},
  number={1},
  pages={96--103},
  year={2023},
  publisher={Royal Society of Chemistry}
}

@article{do2016hydrogenation,
  title={Hydrogenation of the liquid organic hydrogen carrier compound dibenzyltoluene--reaction pathway determination by 1 H NMR spectroscopy},
  author={Do, G and Preuster, P and Aslam, R and B{\"o}smann, A and M{\"u}ller, K and Arlt, W and Wasserscheid, P},
  journal={Reaction chemistry \& engineering},
  volume={1},
  number={3},
  pages={313--320},
  year={2016},
  publisher={Royal Society of Chemistry}
}

@article{wakao1964diffusion,
  title={Diffusion and reaction in porous catalysts},
  author={Wakao, Noriaki and Smith, JM},
  journal={Industrial \& Engineering Chemistry Fundamentals},
  volume={3},
  number={2},
  pages={123--127},
  year={1964},
  publisher={ACS Publications}
}

@article{geisselbrecht2024modeling,
  title={Modeling of the Continuous Dehydrogenation of Perhydro-Dibenzyltoluene in a Cuboid Reactor},
  author={Gei{\ss}elbrecht, Michael and Benker, Rachid and Seidel, Alexander and Preuster, Patrick},
  journal={Energy Technology},
  pages={2300813},
  year={2024},
  publisher={Wiley Online Library}
}

@article{qiu2017upscaling,
  title={Upscaling multicomponent transport in porous media with a linear reversible heterogeneous reaction},
  author={Qiu, Ting and Wang, Qinglian and Yang, Chen},
  journal={Chemical Engineering Science},
  volume={171},
  pages={100--116},
  year={2017},
  publisher={Elsevier}
}

@article{park2021experimental,
  title={Experimental assessment of perhydro-dibenzyltoluene dehydrogenation reaction kinetics in a continuous flow system for stable hydrogen supply},
  author={Park, Sanghyoun and Naseem, Mujahid and Lee, Sangyong},
  journal={Materials},
  volume={14},
  number={24},
  pages={7613},
  year={2021},
  publisher={MDPI}
}

@article{van2024intensified,
  title={Intensified swirling reactor for the dehydrogenation of LOHC},
  author={Van Hoecke, Laurens and Kummamuru, Nithin B and Pourfallah, Hesam and Verbruggen, Sammy W and Perreault, Patrice},
  journal={International Journal of Hydrogen Energy},
  volume={51},
  pages={611--623},
  year={2024},
  publisher={Elsevier}
}

@article{huang1965mathematical,
  title={Mathematical models for mass transfer accompanied by reversible chemical reaction},
  author={Huang, Chen-Jung and Kuo, Chiang-Hai},
  journal={AIChE Journal},
  volume={11},
  number={5},
  pages={901--910},
  year={1965},
  publisher={Wiley Online Library}
}

@article{rios2023revisiting,
  title={Revisiting Isothermal Effectiveness Factor Equations for Reversible Reactions},
  author={Rios, William Q and Antunes, Bruno and Rodrigues, Al{\'\i}rio E and Portugal, In{\^e}s and Silva, Carlos M},
  journal={Catalysts},
  volume={13},
  number={5},
  pages={889},
  year={2023},
  publisher={MDPI}
}

@Article{GreenH2Review2023,
  author       = {Ivanova, M. E. and Peters, R. and Müller, M. and Haas, S. and Seidler, F. and Mutschke, G. and Eckert, K. and Röse, P. and Calnan, S. and Bagacki, R. and Schlatmann, R. and Grosselindemann, C. and Schäfer, L.-A. and Menzler, N. H. and Weber, A. and van de Krol, R. and Liang, F. and Abdi, F.F. and Brendelberger, S. and Neumann, N. and Grobbel, J. and Roeb, M. and Sattler, C. and Duran, I. and Dietrich, B. and Hofberger, C. and Stoppel, L. and Uhlenbruck, N. and Wetzel, T. and Rauner, D. and Hecimovic, A. and Fantz, U. and Kulyk, N. and Harting, J.  and Guillon, O.},
  title        = {Technological Pathways to Produce Compressed and Highly Pure Hydrogen from Solar Power},
  journal      = {Angewandte Chemie International Edition},
  volume       = {2023},
  pages        = {e202218850},
  year         = {2023},
  doi          = {10.1002/anie.202218850}
}

@article{fan2025continuous,
  title={Continuous dehydrogenation of dodecahydro-N-ethylcarbazole for hydrogen production in a micro-packed bed reactor},
  author={Fan, Yiwei and Xu, Yanlin and Wang, Peixia and Liu, Wei and Zhang, Jisong},
  journal={International Journal of Hydrogen Energy},
  volume={103},
  pages={787--796},
  year={2025},
  publisher={Elsevier}
}

@article{kadar2024boosting,
  title={Boosting power density of hydrogen release from LOHC systems by an inverted fixed-bed reactor design},
  author={Kadar, J and Gackstatter, F and Ortner, F and Wagner, L and Willer, M and Preuster, Patrick and Wasserscheid, P and Gei{\ss}elbrecht, M},
  journal={International Journal of Hydrogen Energy},
  volume={59},
  pages={1376--1387},
  year={2024},
  publisher={Elsevier}
}

@article{gambini2024flow,
  title={Flow rate control in a plug-flow reactor for liquid organic hydrogen carriers dehydrogenation},
  author={Gambini, Marco and Guarnaccia, Federica and Manno, Michele and Vellini, Michela},
  journal={International Journal of Hydrogen Energy},
  volume={62},
  pages={375--388},
  year={2024},
  publisher={Elsevier}
}

@article{park2023kinetic,
  title={Kinetic analysis of dibenzyltoluene hydrogenation on commercial Ru/Al2O3 catalyst for liquid organic hydrogen carrier},
  author={Park, Sanghyoun and Abdullah, Malik Muhamamd and Seong, Kwanjae and Lee, Sangyong},
  journal={Chemical Engineering Journal},
  volume={474},
  pages={145743},
  year={2023},
  publisher={Elsevier}
}

@article{auer2025enhancing,
  title={Enhancing the power density of hydrogen release from LOHC systems by high Pt loadings on hierarchical alumina support structures},
  author={Auer, Franziska and Solymosi, Thomas and Erhardt, Chris and Collados, Carlos Cuadrado and Thommes, Matthias and Wasserscheid, Peter},
  journal={International Journal of Hydrogen Energy},
  volume={100},
  pages={1282--1290},
  year={2025},
  publisher={Elsevier}
}

@article{wang2021capillary,
  title={Capillary equilibrium of bubbles in porous media},
  author={Wang, Chuanxi and Mehmani, Yashar and Xu, Ke},
  journal={Proceedings of the National Academy of Sciences},
  volume={118},
  number={17},
  pages={e2024069118},
  year={2021},
  publisher={National Academy of Sciences}
}
\end{document}